\documentclass[aps,prb,twocolumn,superscriptaddress,showpacs,longbibliography]{revtex4-1}

\usepackage{amsmath}
\usepackage{amsfonts}
\usepackage{amssymb}
\usepackage{graphicx}
\usepackage{color}
\usepackage{bm}
\usepackage{hyperref}
\usepackage{mathtools}
\usepackage{bbold}
\usepackage{empheq}
\usepackage[english]{babel}
\usepackage{pgfgantt}
\usepackage{dsfont}
\usepackage{enumerate}
\usepackage{comment}
\usepackage{soul}

\newcommand{\bk}{\mathbf{k}}
\newcommand{\bn}{\mathbf{n}}
\newcommand{\bx}{\mathbf{x}}
\newcommand{\be}{\mathbf{e}}

\newcommand{\bv}{\mathbf{v}}

\newcommand{\by}{\mathbf{y}}
\newcommand{\tr}{\mathrm{tr}}

\newcommand{\pcsadd}{Center for Theoretical Physics of Complex Systems, Institute for Basic Science (IBS), Daejeon 34126, Korea}

\begin{document}

\title{Quantum Hall criticality in Floquet topological insulators}

\author{Kun Woo Kim}
\affiliation{Institut f\"ur Theoretische Physik, Universit\"at zu K\"oln,
Z\"ulpicher Stra\ss e 77, 50937 K\"oln, Germany}
\affiliation{School of Physics, Korea Institute for Advanced Study, Seoul 02455,  Korea}
\affiliation{\pcsadd}

\author{Dmitry Bagrets}
\affiliation{Institut f\"ur Theoretische Physik, Universit\"at zu K\"oln,
Z\"ulpicher Stra\ss e 77, 50937 K\"oln, Germany}

\author{Tobias Micklitz}
\affiliation{ 
Centro Brasileiro de Pesquisas F\'isicas, Rua Xavier Sigaud 150, 22290-180, Rio de Janeiro, Brazil}

\author{Alexander Altland}
\affiliation{Institut f\"ur Theoretische Physik, Universit\"at zu K\"oln,
Z\"ulpicher Stra\ss e 77, 50937 K\"oln, Germany}

\begin{abstract}
The anomalous Floquet Anderson insulator (AFAI) is a two dimensional periodically
driven system in which  static disorder stabilizes two topologically distinct phases
in the thermodynamic limit. The presence of a unit-conducting chiral edge mode and
the essential role of disorder induced  localization are reminiscent of the  integer
quantum Hall (IQH) effect. At the same time, chirality in the AFAI is introduced via
an orchestrated driving protocol, there is no magnetic field, no energy conservation, and no (Landau level)
band structure. In this paper we show that in spite of these differences the AFAI
topological phase transition is in the IQH universality class. We do so by
mapping the system onto an effective theory describing phase coherent transport in
the system at large length scales. Unlike with other disordered systems, the form of
this theory is almost fully determined by symmetry and topological consistency
criteria, and can even be guessed without calculation. (However, we back this
expectation by a first principle derivation.) Its equivalence to the Pruisken theory
of the IQH demonstrates the above equivalence. At the same time it makes predictions
on the emergent quantization of transport coefficients, and the delocalization of
bulk states at quantum criticality which we test against numerical simulations.
\end{abstract}


\maketitle

\section{Introduction}
\label{sec:Introduction}
Topological Floquet insulators (TFI) are gapped quantum systems with topological
structures generated by a periodic drive. The material class includes conventional
topological insulators subject to external driving~\cite{oka2009,lindner2011floquet},  quantum walks~\cite{Kitagawa:2012,edge2015localization}, periodically
kicked matter~\cite{dahlhaus2011quantum,tian2016}, driven  Anderson insulators\cite{titum2016}, and others~\cite{jiang2011majorana, rechtsman2013photonic,potirniche2017floquet, higashikawa2019floquet}. TFIs share many observable properties with static topological insulators. Specifically, they feature phases with quantized edge modes which are protected by a gapped bulk and separated from trivial phases via gap closing transitions. However, these analogies notwithstanding, the physics of topological protection in Floquet quantum matter is based on principles quite different from those in static systems\cite{kitagawa2010topological, rudner2013,roy2017periodic}.  

To see why, recall that the topological invariants of a static topological insulator
describe twists in its free fermion ground state~\cite{Kitaev2009}. The identity of the latter is
protected by a spectral gap, or an excitation gap, in the presence of disorder.
However, absent energy conservation,  topological order in a
Floquet system cannot be based on features of individual of its (quasi--)energy
bands. Any topological classification must address the totality of \emph{all} Floquet
eigenstates, which is information equivalent to the Floquet operator, $\hat U$,
itself.  For example, in systems void of time-reversal and/or charge-conjugation
symmetry (class A), the Floquet operator $\hat U \in \mathrm{U}(N)$ is just an
$N$-dimensional unitary matrix, and topological invariants must be described via
those supported by $\mathrm{U}(N)$. This brings about the seemingly paradoxical situation that there do
exist class A TFIs in two-dimensional space~\cite{rudner2013}, while the unitary group in two
dimensions is topologically empty. (Invariants for the unitary group exist only in odd--dimensional parameter spaces.)

It has been understood~\cite{titum2016} that the resolution of this conundrum lies in the
stabilization of topological phases by \emph{disorder}. At first sight, this may seem counterintuitive: the addition of weak disorder to a clean system removes band gaps and leads to the hybridization of surface states with spatially extended bulk states (According to Mott's criterion energetically degenerate localized and extended states cannot coexist) compromising the identity of the latter. However, the situation changes in the thermodynamic limit, where disorder induced localization renders the bulk fully insulating, which leads to surface state protection in full analogy to the conspiracy of localization and topology in the integer quantum Hall (IQH) effect. This
argument indicates that topological invariants in topological Floquet matter are
\emph{emergent invariants} which become well defined only in the thermodynamic limit. Intuitively,
the scaling parameter controlling the size of the system  provides the  third
parameter required to define an invariant for $\mathrm{U}(N)$.

As a corollary, the \emph{phase transitions} between different topological sectors
must be Anderson localization/delocalization transitions. In static two-dimensional
systems lacking symmetries besides unitarity we know only one topological Anderson
transition, the IQH transition. This raises the question for the universality class
of the TFI phase transition in two-dimensions, is it in the IQH class, or not?
Arguments in either direction may be put forward. On the one hand, it seems natural
that an Anderson transition between phases with different chiral edge modes should be
intimately related to the IQH transition. On the other hand, the absence of energy
conservation and of a bulk Landau level structure (with delocalized Landau level
centers deep below the Fermi surface) point in a different direction.

All these questions can be addressed  on the example of the AFAI, a beautiful paradigm of
two-dimensional topological Floquet quantum matter introduced in Ref.~\cite{rudner2013}. In this
paper, we will present a first principle theory of this TFI subject to maximally strong disorder. We will provide analytical evidence backed by numerical observation that the system is in the IQH universality class. The general strategy of our approach, a mapping of the microscopic theory to an effective theory describing its physics at large length and time scales,   
carries over to other forms of topological Floquet quantum matter, both insulating and
gapless. A central message of this approach is that, unlike in the
physics of topological band insulators, the presence of 
disorder is key to the stabilization of topological phases in Floquet matter. The ensuing
ensemble averaged theories, are simple and depend only on few (two in general) system
parameters characterizing the interplay of bulk localization and topology. We will discuss how the flow of these couplings reflects in microscopic observables and apply exact diagonalization to probe the critical regime and compare to the predictions of the effective theory.

The rest of the paper is organized as follows. After a brief review of the AFAI in section~\ref{sec:AFAI} we discuss its effective theory in section~\ref{sec:effectiveTheory}. As stated above, this will be done on the basis of consistency reasoning (no previous knowledge in the field theory of disordered systems is required to follow this construction). In section~\ref{sec:topological_response_theory} we discuss the connection between the effective theory and observables and put this connection to a numerical test. Finally, in section~\ref{sec:AFAIFieldTheory} (optional reading) we show how the effective theory is derived from first principles.
Our technical calculations are relegated to few Appendices. 

\section{AFAI}
\label{sec:AFAI}

\begin{figure}
	\includegraphics[width=8cm]{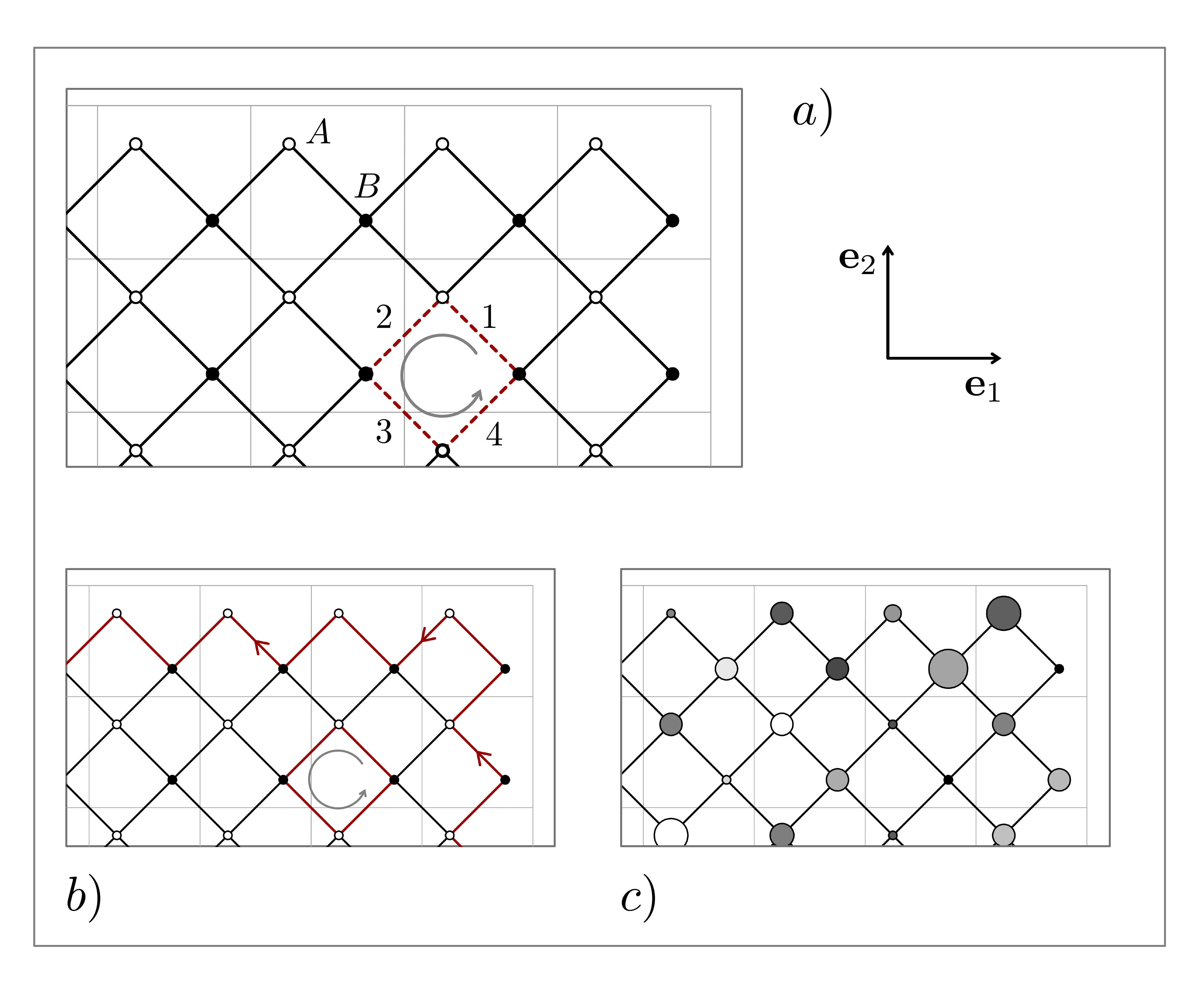}
	\includegraphics[width=7.5cm]{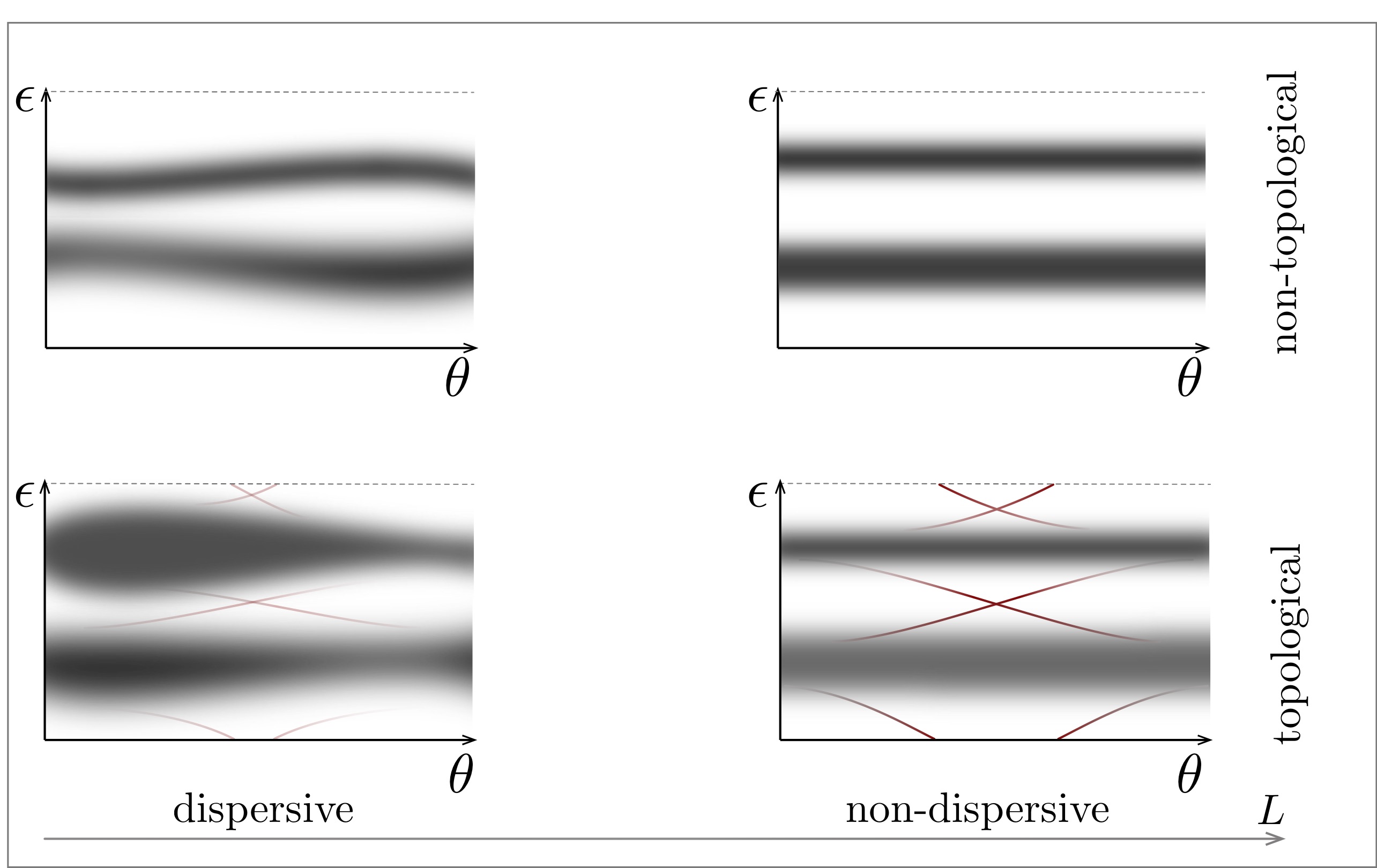}
	\caption{Top: a) four step driving protocol for sequential hopping around the
	plaquettes of a square lattice with tunable amplitude, $t$. 
b) Edge mode in the limit of unit
	probability hopping. c) Application of site diagonal disorder in the initial step. 
    Bottom: schematic of the sensitivity of the Floquet quasi energy
	spectrum to changes in boundary conditions. Left: for finite system sizes the
	spectrum contains quasi-energy bands, whose dispersiveness diminishes upon
	increasing system size. In a pre-topological regime, the precursors of a chiral
	mode are formed. Right: in the thermodynamic limit, bulk modes have become
	non-dispersive, indicating complete localization. In addition to that, the spectrum contains a robust
	chiral edge mode.}
	\label{fig:AFAI}
\end{figure}

The AFAI is a two dimensional driven disordered  system displaying a topological
phase transition~\cite{rudner2013}. It is defined via a five-step driving protocol on a two-dimensional
square lattice (cf. Fig.~\ref{fig:AFAI} a)):   $U_d\equiv   U U_\phi \equiv U_4 U_3 U_2 U_1 U_\phi$.
Here, the unitary operators $U_{1,2,3,4}$ describe deterministic nearest neighbor directed hopping along the bonds of the lattice, represented in a 45$^\circ$ rotated orientation in Fig.~\ref{fig:AFAI} to conveniently describe the ($A/B$) unit cell structure crucial to the definition of the theory. In momentum space these four operators read
\begin{align}
	\label{eq:FloquetDef}
	U_i(\bk) & =\exp\bigl(i t V_i(\bk)\bigr)=\cos(t)+i\sin(t)V_i(\bk), \cr       
	    & V_i(\bk)=\left( \begin{matrix}                     
	    & e^{-i\bk \cdot \bv_i}\cr e^{i\bk \cdot \bv_i} 
	\end{matrix} \right), 
\end{align}
where the matrix structure is in ($A/B$)~--~space and the definition of the hopping vectors, $\bv_i$, 
follows from inspection of the figure as  $(\bv_{1},\bv_{2},\bv_{3},\bv_{4})=(0,-\be_1,-\be_1+\be_2,\be_2)$ 
with $\be_{1,2}$ being lattice unit vectors in the horizontal/vertical direction. The dimensionless parameter  $t$ determines the amplitude of the hopping, interpolating between a stationary limit, $t=0$, and hopping with unit probability, $t=\pi/2$. Finally,
a site diagonal operator
\begin{align}
	U_\phi(\bx)\equiv \left( \begin{matrix}
	e^{i \phi_A (\bx) } & \cr & e^{i\phi_B (\bx)}
	\end{matrix} \right) 
\end{align} 
introduces disorder as random phases at the lattice points. 
We here consider the case of maximal disorder, where $e^{i\phi_{A/B}(\bx)}$ at the lattice sites $\bx = (x_1,x_2)$ 
are independently and identically distributed over the unit circle~(Fig.~\ref{fig:AFAI}, c)).

The presence of topology in the system is understood by inspection of two limiting
cases: for $t=0$ there is no hopping at all, and the clean part of the Floquet
operator, $U\equiv \prod_{i=1}^4 U_i$ reduces to the unit operator. By contrast,
for $t=\pi/2$, we have hopping with unit probability, which means that the bulk
$U$, sends particles back to their point of departure and likewise acts as a
identity operation. However, in this case, the inert bulk is surrounded by an edge
mode encircling the system in counter clockwise direction (Fig.~\ref{fig:AFAI}, b)).

Crucially, however, the two points $t=0,\pi$ do not define topological \emph{phases}
of the clean system. Any perturbation away from these configurations, introduces bulk
extended states, hybridizing the boundaries and compromising their chiral edge mode.
The definition of perturbatively robust phases requires the stabilizing influence of
disorder. Disordered two dimensional systems in class A are subject to Anderson
localization. To anticipate its interplay with a topological phase, we consider  a gedanken experiment
where twisted boundary conditions specified by an angle $\theta$ in, say,
$x$-direction are applied. For a   system of finite size, $L$, generic states will respond to
changes in the boundary conditions, thus rendering the set of quasi-energy levels
dispersive under variations of $\theta$ (Fig.~\ref{fig:AFAI}) left. For $t$ closer to
$\pi/2$ than to $0$, signatures of edge modes, still compromised by bulk
hybridization begin to form (bottom left). In the thermodynamic limit, all bulk bands
have become flat, indicating complete localization. At the same time, two chiral edge
modes, counter propagating at the upper and lower edge have formed. The stabilization of the edge modes and the non-dispersiveness of the bulk are flip sides of the same coin. This scenario
indicates that the essential physics of the system is encoded in the flow of two parameters, a bulk transport coefficient, $g(L)$, and a
topological index $\chi(L)$. For generic bare values $(g_0,\chi_0)$, one expects flow
$(g_0,\chi_0)\stackrel{L\rightarrow \infty}\longrightarrow (0,n)$, where $n=0,1$
depending on $\chi_0$. These variants must be separated by a critical surface,
$g_0,\chi_{\mathrm{crit}}$, where  bulk states remain delocalized.

\section{Effective theory} 
\label{sec:effectiveTheory}

To better understand the physics outline above, an effective theory of the
disordered system is needed. The latter must contain structures defined in crystal momentum space, where
$U=U(\bk)$ describes the clean system, and in real space, where the physics of
localization develops. In this paper, we discuss how such a theory is derived from first principles by methods previously developed for
other problems with random unitary operators~\cite{zirnbauer1996, Altland_2015}.  
This construction  reveals differences and analogies to the
localization theory of the IQH effect, and shows how the flowing coupling constants entering the effective theory are related to response functions  which can be measured, e.g., by numerical experiments.

In this paper, we offer two avenues to the construction of the theory. In this
section, it is `derived' on the basis of symmetry reasoning, and a few references to
general elements of localization theory. The present system is special in that such
reasoning is sufficient to almost fully determine its effective long range theory. In
particular, the  structure of the topological terms governing the disordered system
can be understood in this way. In section~\ref{sec:AFAIFieldTheory}, interested
readers find an alternative derivation of the theory which is more technical and
based on a first principle construction.

We first note that all information required to describe transport and topological properties of the system is contained in the retarded and advanced `resolvents',
\begin{align}
	\label{eq:GDef}
	G^+(\phi) \equiv \frac{1}{1-e^{-\delta}e^{i\phi}U_\mathrm{d}},\quad	G^-(\phi) \equiv \frac{1}{1-e^{-\delta}e^{-i\phi }U^\dagger_\mathrm{d}},
\end{align}
where $\delta \to 0^+$.
For example, in Appendix~\ref{app:source_terms}, we show that the Floquet Hall coefficient probing transverse response is given by the sum of two terms, 
$\widetilde\sigma_H= \sigma^{\rm I} + \sigma^{\rm II}$, with
\begin{equation}
	\label{eq:Sigma_H}
	\sigma^{\rm I} = - \frac {1}{2A } \epsilon_{ij} \left\langle {\rm tr} \left(  v_i G^+ v_j G^-\right) \right\rangle_\phi, \quad   
\sigma^{\rm II} =  \frac{1}{2A} \epsilon_{ij} \, {\rm tr}  ( x_i v_j ),
\end{equation}
where $\langle \dots\rangle_\phi=\int_0^{2\pi}\frac{d\phi}{2\pi}(\dots)$ is a quasi-energy average, $A = L_1 L_2$ is the system 
area, we have defined $G^{\pm}\equiv G^{\pm}(0)$,
and
\begin{align}
	\label{eq:vDef}
	v_i \equiv x_i^U\equiv U x_i U^\dagger -x_i, 
\end{align}
is the Floquet analog of the velocity operator in an Hamiltonian system. (The analogy is seen by writing $U=\exp(i t H)$ and considering the limit of short stroboscopic time: $x^U_i \to i[H,x_i]=\partial_{k_i }H=v_i$~\footnote{
Note that $v_i$ is defined w.r.t. the clean part of the Floquet operator, $U$. Had one start from another definition, 
$v_i^d = U_d x_i {U_d}^\dagger - x_i$, one would see that the random part $U_\phi$ doesn't contribute to $v_i^d$
(indeed,  $[U_\phi, x_i]=0$) and hence one finds $v_i^d \equiv v_i$. }
.) 
Physically, the matrix elements
$G^+_{\mu\nu}(\phi,\by,\bx)\equiv \langle \mu, \by|G^+(\phi)|\nu, \bx\rangle$  
with $\mu,\nu \in \{ A, B\}$ referring to the sublattice structure,
describe the retarded Floquet evolution of states in the system. This is best seen in a
discrete time Fourier transform representation $G^+(t,\by,\bx)= \int
\frac{d\phi}{2\pi}e^{i \phi t}G^+(\phi,\by,\bx)=\langle \by|\bx^+_t\rangle$, where
$|\bx_t^+\rangle \equiv U_\mathrm{d}^t|\bx\rangle$, $t\in \Bbb{N}^+$ is the
stroboscopic evolution of an initial localized state $|\bx\rangle$ and we omitted sublattice indicies for brevity. 
In a similar way, $G^-$ describes the advanced evolution under $U_d^{\dagger t}=U_d^{-t}$. 
The issue with these stroboscopically propagated lattice wave functions  is that they are wildly
oscillatory. It has been a crucial insight of early localization theory~\cite{wegner1979,efetov1980zh,pruisken1982anderson, efetov1983kinetics,pruisken1984}  that an
efficient description of disordered quantum systems should be based on wave function
\emph{bilinears}. The corresponding composite degrees of freedom are the
$Q$-matrices of the non-linear $\sigma$-model. Originally introduced within the framework of static disordered systems, their  applicability extends to the present context, as demonstrated by explicit construction in section~\ref{sec:AFAIFieldTheory}. We here take the alternative route to introduce   these degrees of freedom on the basis of qualitative reasoning: 

We define matrices $Q(\bx)=\{Q^{sa,s'a'}(\bx)\}$  carrying a two fold index structure: the first index pair $s,s'=\pm$
distinguishes between advanced and retarded degrees of freedom such that
\begin{align}
	\label{eq:QMeaning}
	Q^{aa'}(\bx)\sim \left(\begin{matrix}
		|\bx^+\rangle \langle \bx^+| &|\bx^+\rangle \langle \bx^-|\cr 
		|\bx^-\rangle \langle \bx^+| &|\bx^-\rangle \langle \bx^-|  
	\end{matrix}  \right)^{aa'}
\end{align}
represents wave function bilinears of either sense of propagation, retarded or
advanced. The meaning of the subscripts is that the time evolution of the $Q$'s is generated by action of powers of $U_d$ $(U_d^\dagger)$ on $|\bx^+\rangle$, ($|\bx^-\rangle$). Crucially, the `adjoint' action generated by $U_d$ and $U_d^\dagger$ contains contributions where rapid phase oscillations cancel out, which makes $Q$ a candidate for an effective slowly fluctuating variable. For example, in $Q$-language, transport correlation functions are
represented as correlation functions of the structure, $\langle \bx| G^+| \by
\rangle\langle \by |G^-|\bx\rangle\sim \langle Q^{+-}(\by)Q^{-+}(\bx)\rangle_Q $, or
`spectral' correlation functions as $\langle \bx| G^+| \bx \rangle\langle \by
|G^-|\by\rangle \sim \langle Q^{++}(\bx)Q^{--}(\by)\rangle_Q$, where the averages $\langle \dots \rangle_Q$ refer to a $Q$-functional average to be discussed momentarily. The second index, $a,a'$
is required by the disorder average. Depending on the chosen method, this can be a
replica index $a=1\dots, R$, with an analytic continuation $R\to 0$ at the end, or a
`super-index' $a=\mathrm{b,f}$ distinguishing between commuting and anti-commuting
components in a supersymmetric approach~\cite{Efetov-book}. Either way, this structure will not play
an essential role in our present discussion, and we choose a replica representation
for definiteness.

The $Q$-matrices possess a distinct internal structure which on robust grounds is
dictated by unitarity conditions~\cite{wegner1979}. A standard representation incorporating these
principles reads, $Q=T \tau_3 T^{-1}$, where $\tau_3$ is a Pauli matrix in the
representation of Eq.~\eqref{eq:QMeaning}, and $T\in \mathrm{U}(2R)$. 
This structure implies that matrices $T$ commuting with $\tau_3$ are irrelevant, which makes $Q$
an element of the coset space $\mathrm{U}(2R)/\mathrm{U}(R)\times \mathrm{U}(R)$. In the single replica case, $R=1$, this is just a two sphere  $S^2=\mathrm{U}(2)/\mathrm{U}(1)\times \mathrm{U}(1)$.
 For general $R$, the field manifolds are more complicated, but their
geometry, coordinates, etc. still resemble those of generalized spheres.

The theory we are looking for must be described by a simple action for the momentum
space degrees of freedom. $U(\bk)$ and the real space field $Q(\bx)$. Principles entering the
definition of this action include (i) locality, (ii) simplicity (i.e. lowest number of
derivatives), (iii)  symmetry under `canonical' transformations of the coordinates $\bx$ and momenta $\bk$, and (iv) internal
symmetry. The latter principle requires that, e.g., a change of reference frames
$U(\bk)\to U_0 U(\bk) U_0^{-1}$, $Q(\bx)\to T_0 Q(\bx)T_0^{-1}$ should leave the
action invariant. In the following we show that these conditions determine the effective action, up to one
inessential numerical constant.

\noindent \emph{Diffusive action:}
Condition (iv) above excludes the presence of zero derivative terms, and (iii) that of first order derivative terms in an effective action. To lowest, second order, an obvious candidate compatible with (iii) reads $S_{0}[U]\equiv \int d^2k\, \tr(\partial_i U \partial_i U^{-1})$, where $d^2k\equiv dk_1 dk_2/(2\pi)^{2}$ and $\partial_i \equiv\partial_{k_i}$. Notice, however, that this term cannot appear on its own in a valid effective action. The reason is that the (canonical) scale transformation $k_i \to b_i k_i$, $x_i \to b^{-1}_i x_i$ does not leave it invariant. The natural $Q$-partner for pairing reads $S_0[Q]\equiv \int d^2x \, \tr(\partial_i Q \partial_i Q)$, where $d^2x=dx_1 dx_2$, and $\partial_{1,2} \equiv \partial/\partial x_{1,2}$. The product $S_\mathrm{diff}[Q,U]\equiv S_0[U]S_0[Q]$ satisfies all criteria listed above. Seen as a real space action it defines
\begin{align}
	\label{SDiff}
	S_\mathrm{diff}[Q]&\equiv {\sigma^{(0)}_{11}\over 8} \int d^2 x \,\tr(\partial_i Q \partial_i Q), \cr 
	\sigma^{(0)}_{11}&\equiv \frac 12  \int d^2k\, \tr(\partial_1 U \partial_1 U^{-1}).
\end{align}
where, the coupling constant $\sigma^{(0)}_{11}$ defines the mobility of the system at the bare level.

\begin{figure}
	\includegraphics[width=8cm]{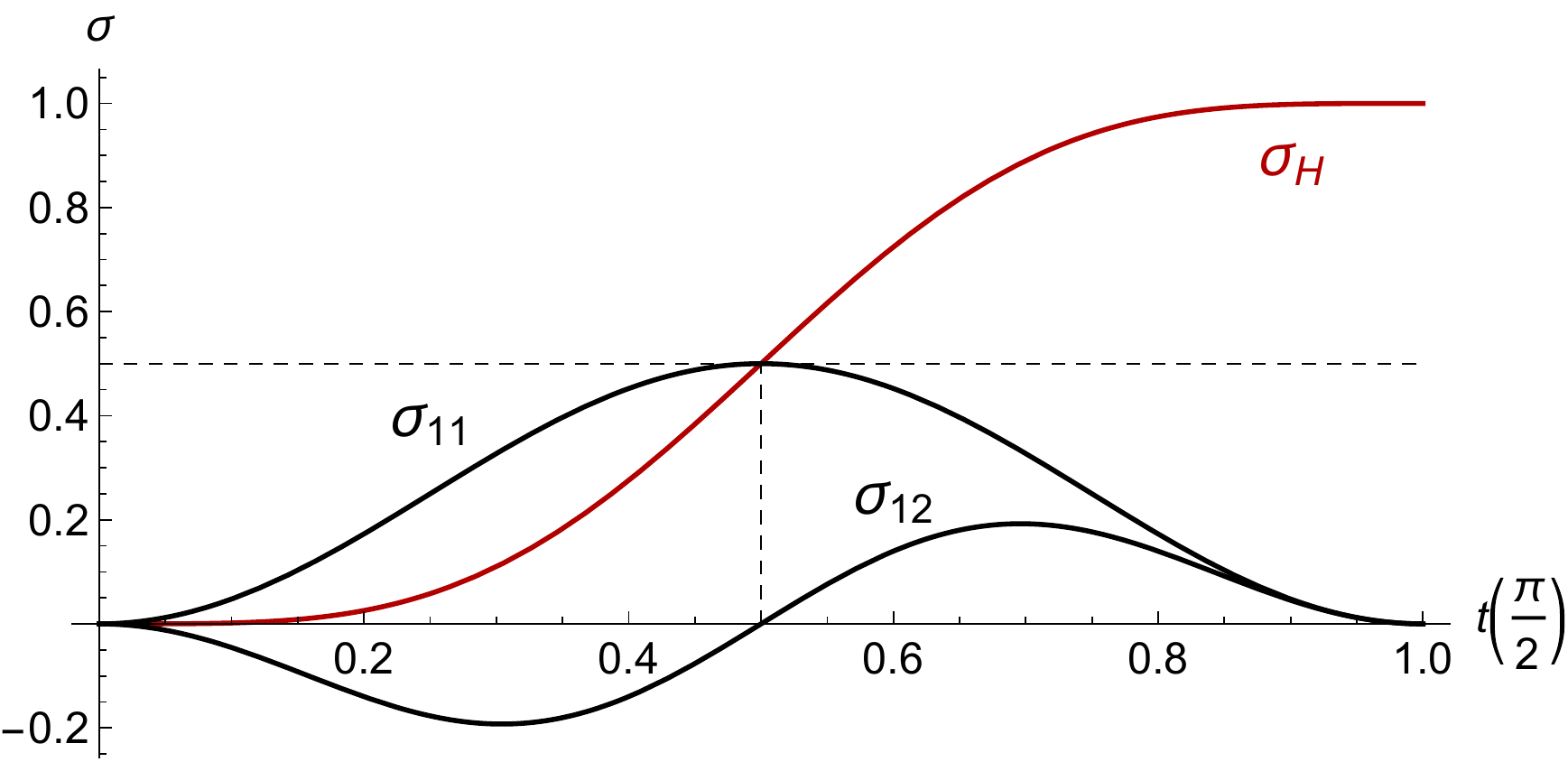}
	\caption{Coupling constants of the nonlinear $\sigma$-model action as a function of $t$ (in units $\pi/2$). 
At $t=\pi/4$ the system passes through a critical point characterized by a half integer quantized value of $\sigma_{H}$ and a maximum of $\sigma_{11}$.}
	\label{fig:sigma}
\end{figure}

Within the $\sigma$-model approach, the conduction properties of disordered metals~\cite{wegner1979} are described by a term of identical structure. In that context, the coupling constant plays the role of the longitudinal conductance (in units of the conductance quantum $\mathrm{e}^2/h$), which motivates the denotation $\sigma_{11}$. 
In the present context, a straightforward calculation yields $\sigma^{(0)}_{11}(t)=\tfrac 12 \sin^2 2t$ (see Fig.~\ref{fig:sigma}). As expected, the mobility approaches zero in the limiting cases of none and unit probability hopping around the lattice plaquettes, respectively. At $t=\pi/4$, a maximum value $\sigma_{11}^{(0)}=1/2$, corresponding to a half of the conductance quantum 
is reached.\footnote{  
For completeness, we mention that the microscopic derivation below leads to the  more general expression 
\begin{align}
	\label{eq:SDiffComplete}
 	S_\mathrm{diff}[Q]\equiv \frac 18 \sum_{ij}\sigma^{(0)}_{ij} \int d^2 x \,\tr(\partial_i Q \partial_j Q),
 \end{align} 
 where $\sigma_{ij}^{(0)}$ plays a role of the symmetric part of the conductivity tensor $\sigma_{11}^{(0)}=\sigma_{22}^{(0)}$, $\sigma_{12}^{(0)}=\sigma_{21}^{(0)}=-\frac{1}{2}\sin^2 2t \cos 2t$. In the present system, without external magnetic field, we do not have an Onsager reciprocity relation ($\sigma_{ij}(H)=\sigma_{ji}(-H)$) and the symmetric contribution to the conductivity tensor may have a skew contribution. However, the bare coupling of this contribution is small and it vanishes at first order in perturbation theory due to the spatial asymmetry of $\int \tr(\partial_1 Q\partial_2 Q)$. We therefore ignore this contribution throughout.}

\noindent \emph{Topological action:} In addition to principles (i-iv), a topological
action must satisfy (v), independence of a metric. In combination with the other principles above this means
that both the real space and the momentum space topological action must be
insensitive to coordinate transformations. Beginning with the real space sector, we
temporarily assume periodic boundary conditions, such that $\bx\in T^2$ becomes the
coordinate of a 2-torus. For $R=1$, $Q$ lives on a sphere, and $Q(\bx)$ defines a map
from the torus to the sphere which can be classified by winding numbers. For example,
using the parameterization, $Q=n_a \tau_a$ with $|\bn|=1$, this invariant is obtained
by the familiar surface integral  $\frac{1}{8\pi} \epsilon^{abc} \int d^2x
\epsilon^{ij} n_a
\partial_i n_b  \partial_j n_c\equiv W\in \Bbb{Z}$. The generalization of this
expression to arbitrary $R$ reads $i S_{\theta}[Q]\equiv \frac{1}{16\pi} \int d^2 x\,
\epsilon^{ij}\,\tr(Q \partial_i Q\partial_j Q)$. This integral determines the
homotopy class of the map $T^2\to \mathrm{U}(2R)/\mathrm{U}(R)\times \mathrm{U}(R),
\bx \mapsto Q(\bx)$.

Turning to the conjugate degree of freedom  $U(\bk)$, the unique topological action definable over a two-dimensional parameter space is the Wess-Zumino-Witten (WZW) term
\begin{align}	
	\label{eq:WZWFunctional}
	\Gamma[U]\equiv -\frac{1}{12\pi} \int d^3k\, \epsilon^{\mu\nu\sigma}\tr(U^{-1}\partial_\mu UU^{-1}\partial_\nu UU^{-1}\partial_\sigma U).
\end{align}
Here, $k=\{k_\mu\}\equiv (s,k_1,k_2)$ is a three-dimensional vector, containing a
parameter $s$ as the zeroth component. We define $U(k)\equiv U(s,\bk)$ as a matrix
obtained by generalization $t\to s$ in Eq.~\eqref{eq:FloquetDef}. In this way,
$U(0,\bk)=\mathds{1}$, and $U(t,\bk)=U(\bk)$, i.e. $s$ is a homotopy parameter, such
that $U(s,\bk)$ interpolates between the unit matrix and $U(\bk)$. The $s$-integration in the definition of $\Gamma[U]$ extends over the interpolation interval, $\int_0^t ds$. 

Geometrically, $\Gamma[U]$ is the three dimensional volume swept out by the map
$U(s,\bk)$ in $\mathrm{U}(2)$, where the latter is defined to cover the interpolation
between $\mathds{1}$ and $U(\bk)$. However, there is an ambiguity in this
construction~\cite{Altland_Simons-book}: instead of interpolating to $\mathds{1}$, one could have chosen
$-\mathds{1}$ as the anchor point. This would have led to a different functional
$\tilde \Gamma[U]$.  This choice, which has a status similar to a gauge ambiguity,
must be physically inconsequential. As with the physically equivalent Dirac spin
quantization condition, the solution is to require that $\Gamma[U]$ couples to the
action via a coupling constant, $\exp(i \kappa \Gamma[U])$ such that $\exp(i \kappa
(\Gamma[U]-\tilde \Gamma[U]))=1$. By construction, $\Gamma[U]-\tilde \Gamma[U]$
equals the integral over the full unitary group, which in the present normalization
equals $2\pi$. This argument shows that the coupling constants, $\kappa$, by which a
WZW term enters the action must be integer quantized.

Above we have seen, that the topological action $i S_{\theta}[Q] \equiv W$ \emph{is} integer quantized. This makes the hybrid  $S_{\mathrm{top}}[Q,U]\equiv \Gamma[U] S_{\theta}[Q]$ the unique topological action consistent with all criteria, (i-v). Emphasizing the real space part, this reads
\begin{align}
		\label{Stop}
	S_\mathrm{top}[Q]&\equiv \frac{\sigma^{(0)}_{H}}{8} \int d^2 x\,\epsilon^{ij}\,\tr(Q \partial_i Q\partial_j Q) , \cr 
	\sigma^{(0)}_{H}&\equiv  \frac{1}{2\pi}\Gamma[U].
\end{align}
Here, $\sigma^{(0)}_H$ is the antisymmetric contribution to the Hall conductivity, labeled by the subscript `$H$' to distinguish it from the symmetric $\sigma_{12}$ defined above. 
For the Floquet of Eq.~\eqref{eq:FloquetDef}, a straightforward computation shows that
$\Gamma[U]=2\pi (1 + 2\cos^2 t) \sin^4 t$. Specifically,  
at $t\equiv t_\mathrm{crit}=\pi/4$, $\sigma^{(0)}_H=1/2$. As we are going to discuss next, this marks the quantum phase
transition of the system.

\noindent \emph{Quantum Hall criticality:} The action
$S[Q]=S_\mathrm{diff}[Q]+S_\mathrm{top}[Q]$ equals the Pruisken action~\cite{pruisken1984} for the IQH.
We need to caution, though, that the derivation of the Pruisken action of the IQH is
parametrically controlled by the assumption of weak disorder (meaning a large
dimensionless product of Fermi energy and scattering time), which translates to large
values of the longitudinal coupling $\sigma^{(0)}_{11}$. In the present problem, the bare
value $\sigma^{(0)}_{11}=\mathcal{O}(1)$ means that the model is derived in a  strong coupling
regime, where the  fields $Q$ strongly fluctuate, and the limitation  to
terms with two derivatives is no longer parametrically controlled.  While higher
order terms are renormalization group irrelevant, the assumption of stability of the
action is not quite innocent, as we know that the IQH critical point itself is
described by a conformal field theory different from the Pruisken model and whose identity is the subject of ongoing
research~\cite{zirnbauer2019integer,bondesan2017gaussian}. Strictly speaking, it is a matter of believe, backed by the numerical evidence presented in subsection~\ref{sub:numerical_test}, that the action $S[Q]$ correctly
describes the system, including at strong disorder.~\footnote{ 
One may consider Floquet models with $N \gg 1$ sites per unit cell, governed by a clean unitary operator $U(\bk) \in \mathrm{U}(N)$. In this case the bare value of the longitudinal conductivity 
is large, $\sigma_{11}^{0} \propto N$,  and the derivation of the Pruisken action becomes parametrically
controlled. However, at large distance scales, we expect this generalization to lie in the same universality class as the $N=1$ system studied here.}
(For completeness, we mention
that for weak disorder the identity of individual of the Chern quasi--energy bands of
the clean AFAI remains visible~\cite{rudner2019floquet}. This may well lead to more complicated physics,
beyond the scope of the present analysis.)

A renormalization group approach to the integration over the $Q$-degrees of freedom~\cite{LevineI,LevineII, LevineIII} 
shows that for off critical values $t\not= t_\mathrm{crit}$, the mobility coefficient $\sigma_{11}$
flows to zero at large length scales. At the same time, the topological angle
$\sigma_{H}$ flows to $0$ or $2\pi$, depending on whether the bare value of $t$ is
smaller or larger than $t_\mathrm{crit}$, respectively. This flow implies that
topological quantization is an emergent feature in the present context. It is
emergent in the sense that at large scales a system governed by a generic bare
Floquet operator becomes RG equivalent to one with $t=0$ or $t=\pi/2$. The continuous
interpolation of this effective Floquet with $U=\mathds{1}$ then does not cover the
unitary group, or it covers it once. The  theory also predicts that 
$\sigma_H=1/2$ defines a critical surface between the two phases. On this surface, the system flows towards the IQH critical point at $\sigma_{11}=\mathcal{O}(1)$, and a conformal fixed point theory, as mentioned above.  

\noindent\emph{Frequency action:} Before concluding this section we note that one
occasionally needs to monitor Green functions of different quasi energy argument,
${\omega}$. Both, differences in ${\omega}$, and the infinitesimal regulator $\delta$ in Eq.~\eqref{eq:GDef} break the rotational symmetry under $\mathrm{U}(2R)$ between advanced and retarded degrees of freedom down to $\mathrm{U}(R)\times \mathrm{U}(R)$. The simplest contribution to the action consistent with the general principles (i-iv) reads
\begin{align}
	\label{eq:FreqAction}
	S_{\omega}[Q]\equiv -i \frac{\omega^+}{4}\int d^2 x\,\tr(Q\tau_3),
\end{align}
where ${\omega}^+\equiv {\omega}+2i\delta$,  ${\omega}$ is the difference between the quasi-energy argument of the retarded and the advanced Green function, and the numerical factor $1/4$ is obtained by the microscopic derivation of section~\ref{sec:AFAIFieldTheory}.

\section{Topological response theory} 
\label{sec:topological_response_theory}

In this section, we discuss the linear response approach relating the couplings of the effective theory to microscopically defined transport correlation functions. Our focus will be on quantities carrying topological
significance such as transverse transport coefficients, edge currents, or bulk
magnetization~\cite{nath2017quan}. 
In the first part of the section, we show how the different observables are read out from the field theory. Emphasis will be put on gauge symmetries establishing non-obvious connections between physically distinct representations and observables. The second part of the section defines these quantities  in terms of the Green functions~\eqref{eq:GDef}. We apply spectral decompositions to translate these  representations to concrete expressions in terms of eigenfunctions and -- values.  
In the third part, we put the theory to test and discuss numerical results for the Hall and longitudinal conductivities for increasing system sizes. Our findings supporting the existence of  quantum Hall criticality in the system.
The master quantities discussed in this section will be a transverse transport coefficient, 
denoted as $\widetilde\sigma_H$, and the longitudinal one,  $\widetilde\sigma_{11}$. 

\subsection{Response coefficients from field theory} 
\label{sub:response_coefficients_from_field_theory}

Within the field theory framework, the full information on finite size scaling and
the approach to an integer quantized fixed point configuration is encoded in the flow
of the coefficients $(\sigma_{11}(L),\sigma_H(L))$. These quantities can be extracted
from the functional integral via the introduction of suitable source terms. For our
purposes, a convenient choice is defined by the generalization~\cite{pruisken1984}
\begin{align}
	\label{eq:SourceDef}
	Q\to S Q S^{-1},\quad S=\exp(q_1\tau_1 + q_2\tau_2)\otimes P^{(1)},
\end{align}
where $q_i=a_i x_i$, $a_{i}$ are constant
parameters, and $(P^{(1)})^{ab}=\delta^{a1}\delta^{b1}$ is a projector onto the first
replica channel. The sourced action is then defined as
$S_\mathrm{diff}[SQS^{-1}]+S_\mathrm{top}[SQS^{-1}]+S_\omega[Q]$, i.e. substitution
of the twisted configuration in the fluctuation parts of the action. (A global
substitution would be removable via the reverse transformation $S QS^{-1}\to S$ in
the functional integral and have no effect.) The sources are designed such that
two-fold differentiation in $a_{1,2}$ reads out the couplings $\sigma_{11}$ and $\sigma_H$.
Specifically, using that $S^{-1}\partial_i S=a_i \tau_i P^{(1)} +\mathcal{O}(a)$,
it is straightforward to verify that 
\begin{align}
\label{eq:FieldTheorySource}
  	&\frac{1}{2i A}\left.\frac{\partial^2}{\partial a_1 \partial a_2} \right|_{a=0}\mathcal{Z}(a)=\\
  	&\quad =- \frac{\sigma_H^{(0)}}{16i A}\epsilon^{ij}\left\langle \int d^2x\tr([Q,\tau_i P^{(1)}][Q,\tau_j P^{(1)}]Q)\right\rangle_Q.\nonumber 
  \end{align} 
  The r.h.s of this expression should be read within the context of a renormalization
  group procedure: starting from the bare value $\sigma_{H}^{(0)}$, one  integrates
  out $Q$-fluctuations at successively increasing length scales. The procedure is
  continued down to an effective length scale $L$, below which $Q$-fluctuations are
  damped out, e.g. due to the presence of an RG relevant frequency mismatch, $\omega$,
  cf. Eq.~\eqref{eq:FreqAction}. For larger length scales, fluctuations are
  suppressed, $Q\simeq \tau_3$, and the r.h.s.~of~\eqref{eq:FieldTheorySource}
  reduces to $\sigma_H=\sigma_H(L)$.  In a similar fashion one derives that
\begin{equation}
\label{eq:FieldTheorySource1}
  	\!\!\!\!\frac{1}{2A}\left.\frac{\partial^2}{\partial a_1^2} \right|_{a=0}\!\!\!\mathcal{Z}(a)=- \frac{\sigma_{11}^{(0)}}{16 A} \left\langle \int d^2x\tr([Q,\tau_1 P^{(1)}]^2\right\rangle_{\!\!Q}, 
  \end{equation}
and the r.h.s of this relation reduces to $\sigma_{11}=\sigma_{11}(L)$ that as above has to be understood in the RG sense.

In Appendix~\ref{app:source_terms} below we demonstrate the equivalence of the l.h.s. of~Eqs.~(\ref{eq:FieldTheorySource}) 
and (\ref{eq:FieldTheorySource1}) 
with the microscopic formula for the Hall~\eqref{eq:Sigma_H} and longitudinal response~\eqref{eq:Sigma_xx}, respectively. This construction demonstrates that the flowing coupling constant of the field theory describes the microscopic Hall response of the system.

\subsection{Linear response} 
\label{sub:linear_response}
To begin we will analyze in more details the topological response of the system $\widetilde\sigma_H$ 
in terms of the correlation 
functions~(\ref{eq:Sigma_H}).
Much as the structurally similar Streda formula of the IQH, $\tilde\sigma_H$ describes the
topological response of the system as the sum of two contributions,
$\sigma^{\mathrm{I}}$ being a velocity--velocity, or current--current response
function, and $\sigma^{\mathrm{II}}$ a `thermodynamic' quantity, probing the local
magnetization of the system. The physics of $\sigma^{\mathrm{II}}$ in connection with
the AFAI has been discussed previously in Ref.~\cite{rudner2013,nath2017quan}. In the following, we demonstrate
how the sum of the two parts gives access to the topological response of the system
both via exact diagonalization and through an effective equivalence, $\widetilde\sigma_H\sim
\sigma_H(L)$ with the flowing coupling constant of the field theory.

To begin this discussion, we assume a system of eigenstates, $|n\rangle$ with
$U_{\rm d}|n\rangle \equiv e^{i\epsilon_n}|n\rangle$, to represent the first contribution as
\begin{eqnarray}
	\sigma^{\mathrm{I}}&=&-\frac{1}{2A}\epsilon_{ij}\sum_{nm}\int\frac{d\phi}{2\pi}\frac{\langle m|v_i |n\rangle \langle n|v_j|m\rangle}{(1-e^{i\phi^++i\epsilon_n})(1-e^{-i\phi^--i\epsilon_m})} \nonumber \\
\label{eq:sigma_12I_sum}
	&=&-\frac{1}{2A}\epsilon_{ij}\sum_{n\neq m}\frac{\langle m|v_i |n\rangle \langle n|v_j|m\rangle}{1-e^{i(\epsilon_n-\epsilon_m)^+}},
\end{eqnarray}
where $x^\pm \equiv x\pm i\delta$. 
At this stage we note the relation
\begin{equation}
\label{eq:frac}
\frac{1}{(1 - e^{  i\omega^+})} = {\cal P} \left(\frac{1}{1 - e^{i\omega}}\right) + \pi\delta(\omega).
\end{equation}
Because of antisymmetric structure of the sum~(\ref{eq:sigma_12I_sum}) only the principal part contributes, and
we are thus left with
\begin{equation}
\label{eq:sigma_12I_final}
\sigma^{\rm I} = - \frac{i}{2 A} \sum_{m\neq n} 
\langle m |  v_1 |n\rangle \langle n | v_2 | m \rangle \cot \left(\frac{\epsilon_m - \epsilon_n}{2}\right) .
\end{equation}
As to the second contribution, we show in Appendix~\ref{sec:proof_of_equiv} that
$\sigma^{\rm II} = \sigma_H^{(0)}$, or explicitly
\begin{equation}
\label{eq:sigma_II_WZW}
\sigma^{\rm II}  = \frac {1}{2 A}\epsilon_{ij} {\rm tr} (x_i v_j) = \frac{1}{2\pi} \Gamma[U],
\end{equation}
with WZW term defined by Eq.~(\ref{eq:WZWFunctional}). Thus this contribution is disorder independent and for a specific model 
at hand reads $\sigma^{\rm II} =  (1 + 2\cos^2 t) \sin^4 t$. In Appendix~\ref{sec:sigma_Hall}  
we demonstrate that in the localized regime the Floquet Hall coefficient, $\widetilde\sigma_H = \sigma^{\rm I} + \sigma^{\rm II}$, 
is quantized and as expected counts a number of chiral edge modes. In this limit
$\widetilde\sigma_H$ matches the topological invariant of Ref.~\onlinecite{titum2016}.

\begin{figure*}[t]
	\includegraphics[width=8.9cm]{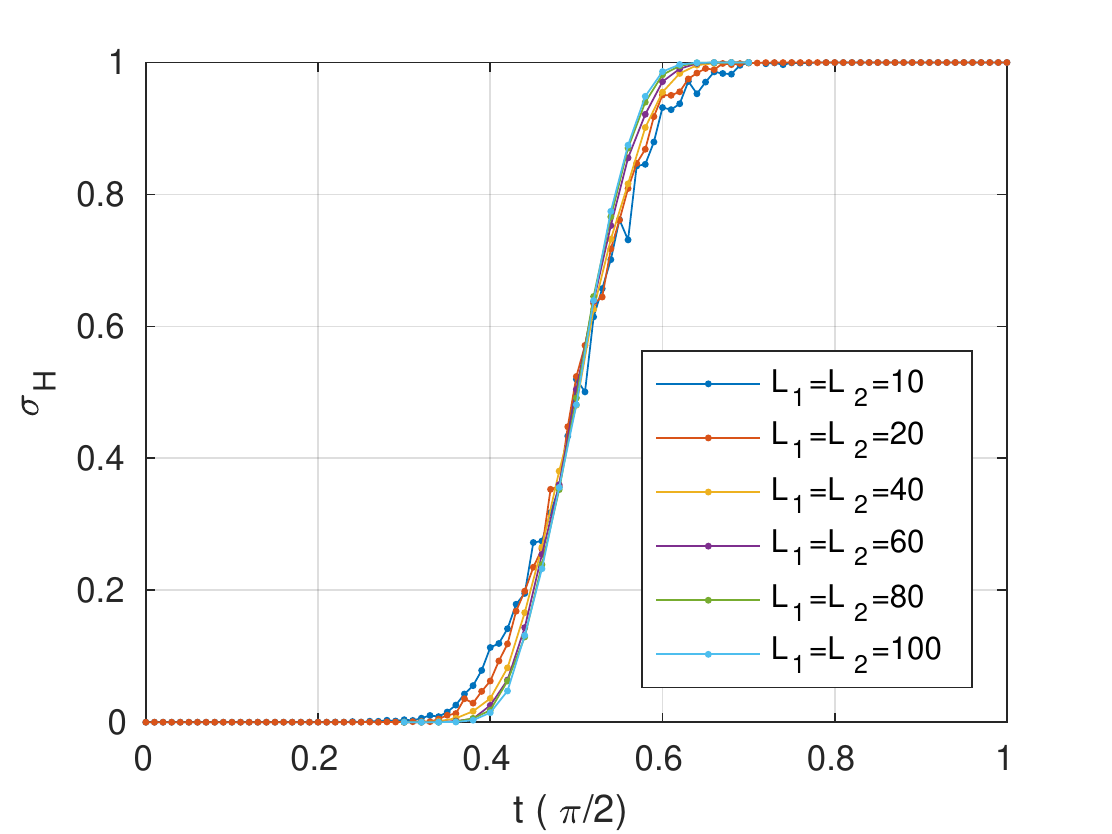}
    \includegraphics[width=8.9cm]{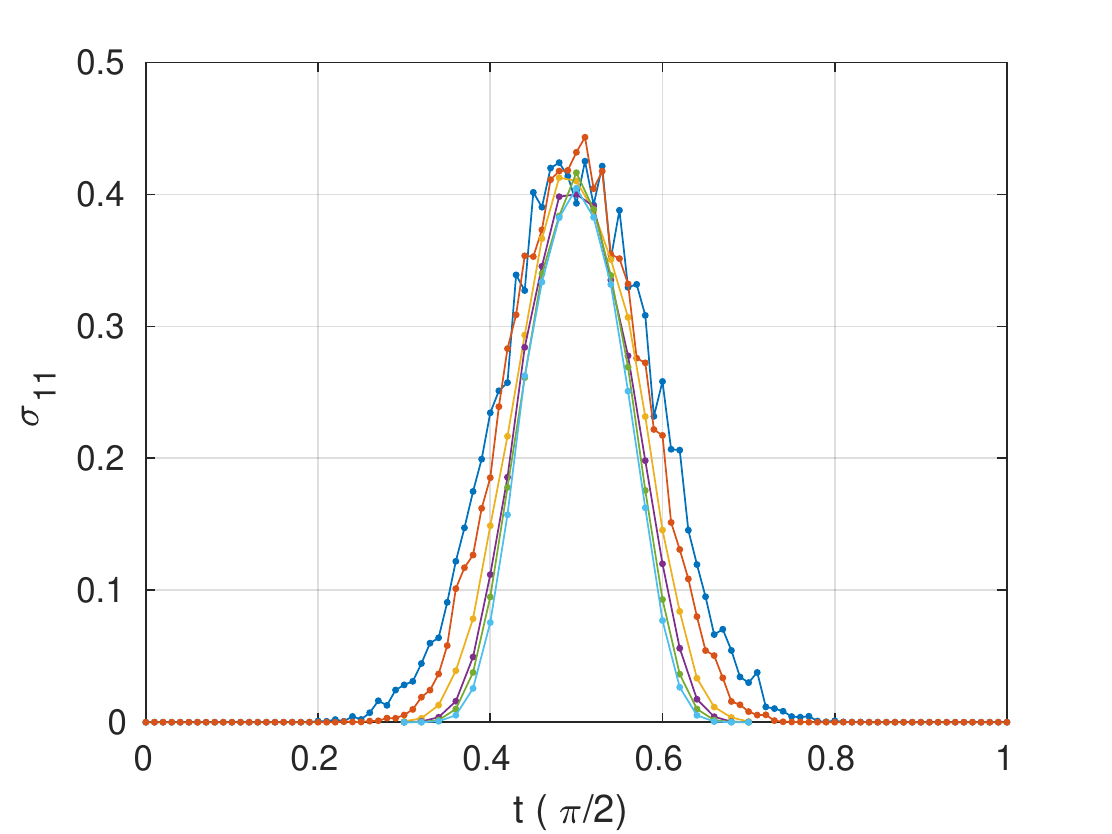}
    \caption{Disorder specific Hall conductivity $\sigma_H$ (left) and longitudinal conductivity $\sigma_{11}$  (right) as a function of $t$    evaluated for different sample sizes $L=10 \div 100$.  At $t=t_{\rm crit}=\pi/4$ the conductivity 
    $\sigma_{11}$ saturates to a critical value $\sigma_{11}^*\simeq 0.4$ at large length scales, 
signaling Anderson de-localization at criticality. The value of  $\sigma_{11}^*$ is close to $\simeq 0.6$ reported
for the critical conductance of conventional IQHE systems, see Ref.~\onlinecite{MirlinEvers:2008} for the review.}
    \label{fig:sigma_num}
\end{figure*}

We now turn to the analysis of the second correlation function,
\begin{equation}
	\label{eq:Sigma_xx}
	\widetilde\sigma_{ii}= \frac {1}{A } \left\langle {\rm tr} \left(  v_i G^+ v_i G^-\right) \right\rangle_\phi 
    -  \frac {1}{2A} {\rm tr} (v_i^2), 
\end{equation}
which defines the longitudinal conductivities and measures the mobility of the system.  Its formal derivation, as well as
the derivation of the Hall response $\widetilde\sigma_H$, see Eq.~(\ref{eq:Sigma_H}), is provided in Appendix~\ref{app:source_terms}.
We note here that the 2nd piece in Eq.~(\ref{eq:Sigma_xx}) plays a role of a diamagnetic term in case of the Floquet system.
Employing as above the spectral decomposition, one then finds
\begin{equation}
\label{eq:sigma_ii_spec}
\widetilde\sigma_{ii} = \frac{1}{A}\sum_{m,n} 
\frac{|\langle m |  v_i |n\rangle|^2}{1-e^{i(\epsilon_n-\epsilon_m)^+}}  - 
\frac{1}{2 A} \sum_{m,n} |\langle m |  v_i |n\rangle|^2. 
\end{equation}
At this point we can use once again the relation~(\ref{eq:frac}). Noting that $\sum_\pm ({1-e^{\pm i \omega}})^{-1} = 1$, one
sees that off-diagonal terms with $n \neq m$ are canceled out in Eq.~(\ref{eq:sigma_ii_spec}). 
We then regularize $\delta(0) = A/2\pi$ to arrive at
\begin{equation}
\label{eq:sigma_ii}
\widetilde\sigma_{ii} = \frac{1}{2 \pi} \sum_{m} |\langle m |  v_i |m\rangle|^2. 
\end{equation}
Results~(\ref{eq:sigma_12I_final}), (\ref{eq:sigma_II_WZW}) and (\ref{eq:sigma_ii}) will serve as a basis for our numerical analysis
discussed in Sec.~\ref{sub:numerical_test} below.

\subsection{Numerical test} 
\label{sub:numerical_test}

\begin{figure*}
   \includegraphics[width=5cm]{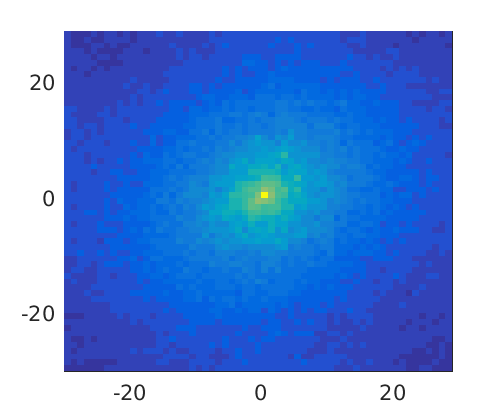}
   \includegraphics[width=5cm]{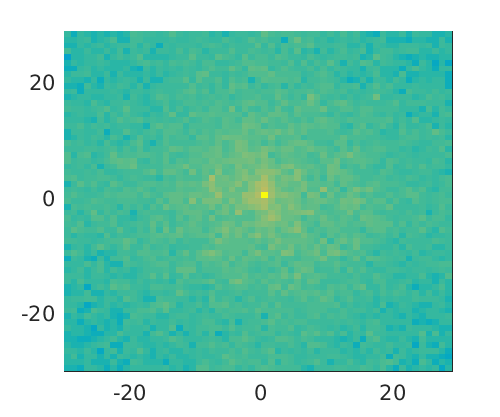}
   \includegraphics[width=5cm]{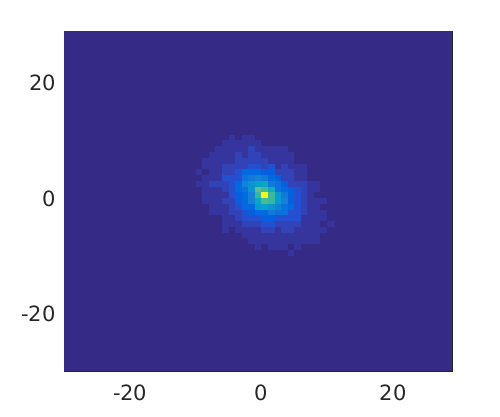}
   \includegraphics[width=.94cm]{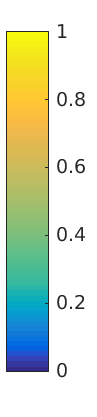}
    \caption{Disorder averaged probability distribution $P(n, {\bf x})$ (see exact definition in the main text) of a wave
    packet to spread on a $60\times 60$ grid after $n=600$ time steps and different values of parameter 
$t = \pi/5 < t_{\rm crit}$ (left), $\pi/4 = t_{\rm crit}$ (middle) and  $7\pi/20 > t_{\rm crit}$ (right). 
The probability distribution is averaged over $100$ disorder realizations, and normalized by the value at $\bf x=(0,0)$. }
    \label{fig:P}
\end{figure*}

In this section, we put the results for longitudinal and Hall conductivities analytically obtained above to a numerical test. 
Our numerical simulations for the disorder specific observables are shown in Fig.~\ref{fig:sigma_num}.  In these plots each $t$ corresponds to a single disorder
realization which varies with $t$. We employ an enlarged unit cells of size $\rm L_1\times L_2$ with twisted boundary conditions to compute the transport coefficients using velocity--velocity response function. For further numerical details, 
see Appendix~\ref{sec:numerics}. One observes that mesoscopic fluctuations which are visible at small system sizes 
($A = 10 \times 10$) are essentially dying out for larger sizes ($A = 10^2 \times 10^2$) where the system becomes self-averaging.
In the limit $L \to \infty$ the Hall conductivity,  $\sigma_H$, tends to the step function $\theta (t-t_{\rm crit})$,
see Fig.~\ref{fig:sigma_num} (left).
At the same time the longitudinal conductivity, $\sigma_{11}$, tends to zero with $L \to \infty$ whenever $t \neq t_{\rm crit}$.  
On the other hand, at $t=t_{\rm crit}$ the system flows to IQH critical point with $\sigma_{11} = {\cal O}(1)$, 
see Fig.~\ref{fig:sigma_num} (right). 
To complement this statement, in Fig.~\ref{fig:P} we show the spreading of the wave packet which
is initially localized at the origin. In more concrete terms the disorder averaged probability
\begin{eqnarray}
P(n, {\bf x}) &=&  \sum_{\mu,\nu \in \{A,B\}} \langle |G_{\mu\nu}^+(n,{\bf x},{\bf 0})|^2 \rangle_\phi \nonumber \\
&\equiv&  \sum_{\mu,\nu \in \{A,B\}} \langle |\langle  \mu, {\bf x} | U^n_{\rm d} | \nu , {\bf 0} \rangle|^2 \rangle_\phi   
\end{eqnarray}
is shown after $n \gg 1$ steps in time. At criticality, $t=t_{\rm crit}$, the wave packet is clearly delocalized. In turn, its degree of
localization is increased as one moves further away from the critical point. 

Transport properties of two- and multi-terminal AFAI setups have been also analyzed numerically recently in Ref.~\onlinecite{Rodriguez:2019}.

\section{Field theory of the AFAI} 
\label{sec:AFAIFieldTheory}

In this section, we derive the field theory discussed in the main text from first
principles. This derivation has been included to make the text self contained and can be skipped by readers not interested in its technical details.

Correlation functions built as products of matrix elements of the resolvent
operators~\eqref{eq:GDef} can be conveniently obtained from the unit-normalized
Gaussian integral representation
\begin{align}
	\label{eq:PsiUFunctional}
1=\mathcal{Z}&=\int D\psi\, \left\langle e^{-\bar \psi \hat{G}^{-1}\psi}\right\rangle.
 \end{align} 
 Here, $\hat{G}=\mathrm{diag}(G^+,G^-)$ is a block diagonal operator containing the
 retarded and advanced Green function on its diagonals, and $\psi=\{\psi_\bx^{sa}\}$ is a vector of Grassmann valued integration variables. The explicit representation of the exponent thus reads
 \begin{align}
 \label{eq:ActionExplicit}
 \bar \psi_\bx^{+a}\left(\delta_{\bx,\by}-e^{i\phi_\bx}U_{\bx\by}\right)\psi_\by^{+a}+ \bar \psi_\bx^{-a}\left(\delta_{\bx,\by}-U^\dagger_{\bx\by}e^{-i\phi_\by}\right)\psi_\by^{-a},		
 \end{align}
where $\exp(i\phi_\bx)$ are the uniformly distributed phases introducing disorder and $U$ is the clean Floquet operator.
Referring to Appendix~\ref{app:source_terms} for details, Green function matrix elements are obtained
from $\mathcal{Z}$ via the introduction of suitable source terms. However, for the
time being, we suppress the presence of these and focus on the unit normalized `partition sum', $\mathcal{Z}$.

\subsection{Color-flavor transformation} 
\label{sub:color_flavor_transformation}


The key to the derivation of an effective disorder averaged theory lies in an
integral identity known as the color-flavor transform (cft)~\cite{zirnbauer1996}. This identity trades
the integration over rapidly fluctuating phases, $\phi_\bx$, for the integration over
a composite field $Z_\bx$, and in this way represents a unitary equivalent of the
Hubbard-Stratonovich transformation for hermitian operators. The general cft identity
reads
\begin{align}
	\label{eq:cftNative}
	&\langle \exp(\bar \psi^{+a}_\bx e^{i\phi_\bx}\varphi^{+a}_\bx+\bar \varphi^{-a}_\bx e^{-i\phi_\bx}\psi^{-a}_\bx)\rangle_\phi=\cr 
	&\qquad =\langle \exp(\bar \psi^{+a}_\bx Z_\bx^{ab} \psi^{-b}_\bx-\bar \varphi^{-a}_\bx Z_\bx^{\dagger ab} \varphi^{+b}_\bx)\rangle_Z.
\end{align}
Here, $\langle \dots\rangle_\phi$ is the average over uniformly distributed phases,
$\prod_\bx \frac{1}{2\pi}\int d\phi_\bx$ with
$\bx  = (\mu, x_1, x_2)$ combining  cite and sublattice indices.  
 On the r.h.s., $Z_{\bx}=\{Z_\bx^{a,b}\}$ are
complex $R\times R$ matrices, and the functional average is defined as $\langle \dots
\rangle_Z=\prod_\bx \int dZ_\bx \,\det(1+Z_\bx Z_\bx^\dagger)^{-1}$. (In this
expression for the measure, the replica limit $R\to 0$ has already been taken. This
simplifies the notation and does not affect any results.) The integral
transform~\eqref{eq:cftNative} is exact and holds for arbitrary Grassmann vectors
$\{\psi_\bx\}$ and $\{\varphi_\bx\}$.

\begin{figure}[b]
	\includegraphics[width=4.5cm]{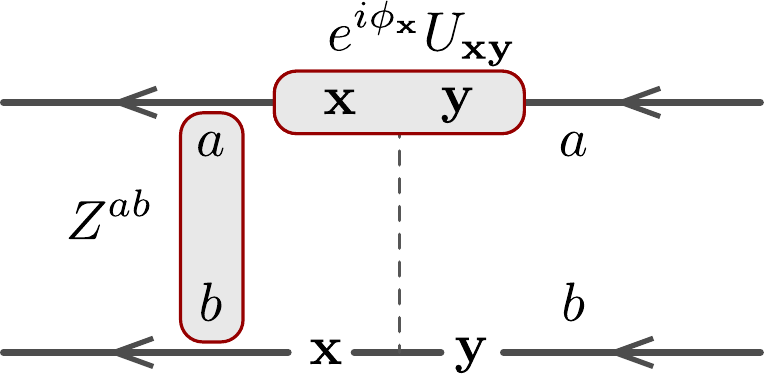}
	\caption{The idea of the color--flavor transformation: individual retarded, $\psi^+$, (top) and advanced, $\psi^-$, (bottom) wave function amplitudes fluctuate rapidly due to phase fluctuations. The composite degree of freedom $Z$ pairs these amplitudes to slowly fluctuating bilinears $\bar \psi^+ \psi^-$.}
	\label{fig:cft}
\end{figure}

The conceptual meaning of the cft is illustrated in Fig.~\ref{fig:cft} for the choice
$\varphi_\bx=U_{\bx\by}\psi_\by$ relevant for our application: due to the presence of
the random phases $e^{i\phi_\bx}$, the bilinears $\bar \psi^a_\bx
e^{i\phi_\bx}U_{\bx\by}\psi^a_\by$ are rapidly fluctuating functions of the `color
index' $\bx$ carrying a conserved `flavor' singlet index $a$. The transformation
trades them for bilinears $\bar \psi^{+a}_\bx Z_{\bx}^{ab}\psi^{-b}_\bx$ and
$(\bar\psi^{-b} U^\dagger)_\bx Z_\bx^{\dagger ba} (U \psi^{+a})_\bx$, which now carry
a `flavor' structure in the replica indices but, for slowly varying $Z$ no longer
fluctuate rapidly in the indices $\by$. (The cft is called `color-flavor-transformation' alluding to a similar setting in QCD, where the role of the fields $\psi$ is taken by quarks, that of $U$ by strong color-interactions, and that of $Z$ by the color-neutral yet flavored slow meson fields.)

Applied to the action~\eqref{eq:ActionExplicit}, the cft leads to the representation
\begin{align}
	\label{eq:CFTAFAI}
	\mathcal{Z}&= \left\langle e^{\bar \psi^+(\mathds{1}-Z)\psi^-+\bar \psi^+(\mathds{1}+U^\dagger Z^\dagger U)\psi^-)}\right\rangle=\cr 
	&=\int DZ \,e^{-\tr\ln(1+ZZ^\dagger)+\tr\ln(1+ZU^\dagger Z^\dagger U)}.
\end{align}
Here, the flavor indices are implicit, and the presence of the convergence generator
$e^{-\delta}$ is suppressed. In the second line, we have performed the Gaussian integral over $\psi$ and elevated the determinantal measure term to
become part of the action. The action now makes the tendency to slow fluctuations in
the $Z$-fields explicit: spatially homogeneous configurations $Z_\bx=\mathrm{const.}$
commute with $U$ and make the two terms in the action cancel. This shows that the
functional integral will be dominated by slowly fluctuating configurations and is
tailored to a gradient expansion.

Before turning to this expansion, we introduce the degrees of freedom $Q=T\tau_3 T^{-1}$ central to the discussion in section~\ref{sec:effectiveTheory}. Defining
\begin{align}
	\label{eq:TDef}
	T\equiv \left(\begin{matrix}
		\mathds{1}& Z\cr -Z^\dagger & \mathds{1}
	\end{matrix}  \right),
\end{align}
the matrices $Z$ acquire a status as linear coordinates of the nonlinear manifold of
$Q$-matrices. Specifically, for $R=1$, where  $Q$ defines a two-sphere, and $Z\equiv
z$ is just a complex number,  an explicit computation of the unit vector
$\mathbf{n}(z,\bar z)$ in $Q=\tau_i n_i$ shows that $z$ is the coordinate of a
stereographic projection, or Riemann sphere representation. For general $R$, the
matrices $Z$ define linear coordinates of the symmetric space
$\mathrm{U}(2R)/\mathrm{U}(R)\times \mathrm{U}(R)$ such that $Z=0$ represents the
`north pole', $Q=\tau_3$, and $Z\to \infty$ the `south pole', $\theta=-\tau_3$.

A straightforward manipulation of block matrices brings the above representation of the action into the form
\begin{eqnarray}
	\label{eq:ProtoTypeAction}
	&&S=-\frac{1}{2}\sum_{s=\pm}\tr \ln \left(1+X_sP^s\right),\\
	&&X_+\equiv T^{-1}U^{\dagger} [T,U] , \quad X_-\equiv T^{-1}U [T,U^\dagger], \nonumber
\end{eqnarray}
where $P^\pm=\{\delta^{s,\pm}\delta^{s',\pm}\}$ are projectors onto the retarded or
advanced index sector.  In this form, the action is tailored for a gradient
expansion. The reason is that the commutators individually vanish for constant $T$.
This means that in a continuum theory an expansion up to $n$th order in these will
contain at least $n$ derivatives. For our purposes an expansion to second order is
sufficient,
\begin{eqnarray}
	&&S=S^{(1)}+S^{(2)},\\
	&&S^{(1)}=-\frac{1}{2}\sum_s \tr(X_s P^s),\quad S^{(2)}=\frac{1}{4}\sum_s \tr((X_s P^s)^2). \nonumber
\end{eqnarray}
The continuum representation of these expressions  is most economically formulated in the language of Wigner transforms. Temporarily using carets to distinguish operators $\hat X(\bx_1,\bx_2)$ from functions, we define
\begin{align*}
	X(\bx,\bk)=\int d^d x\, e^{i\bk\cdot \Delta\bx} \hat X(\bx+\frac{1}{2}\Delta \bx,\bx-\frac{1}{2}\Delta \bx).
\end{align*}
Transformed in this way, the real-space operators $\hat T(\bx)\to T(\bx)$ become
functions of $\bx$ and the momentum-space $\hat U(\bk)\to U(\bk)$ functions of the
conjugate momentum, $\bk$. In Wigner language, matrix products are evaluated as Moyal
products, $(\hat X\hat Y)\to (X\ast Y)=
Xe^{\frac{i}{2}(\partial'_{x_i}\partial_{k_i}-\partial'_{k_i}\partial_{x_i})}Y$, where
derivatives carrying a prime act to the left and we omitted the arguments in
$X(\bx,\bk)$ for brevity. Specifically, for operators diagonal in either the $\bx$--
or the $\bk$--representation, this defines the Moyal expansions
\begin{align}
\label{MoyalDiagonal}
	T * U&=T U+\frac{i}{2} \partial_{i} T \partial_{i} U+\dots,\cr 
	U * T&=U T-\frac{i}{2} \partial_{i} U \partial_{i} T+\dots,
\end{align}
where the  derivatives $\partial_i$ act either on $\bx$ or $\bk$. 
Finally, note that in Moyal language traces are to be interpreted as $\tr(\dots)=\int d^2 x\int d^2k \,\tr(\dots)$, where the measure $d^2k=dk_1 dk_2 /(2\pi)^2$, and the trace on the right extends over the internal indices of the theory. 

\subsection{Diffusive action} 
\label{sub:diffusive_action}

As a warm up to the somewhat more involved
topological action, we consider the derivation of Eq.~\eqref{SDiff}. This action is
obtained from the continuum-Moyal expansion of the second order term $S^{(2)}$. At
this level, it is sufficient to Moyal-expand each of the factors $X_s$ to first order
in derivatives. Application of the rules Eq.~\eqref{MoyalDiagonal} gives
\begin{eqnarray*}
	&&X_{\pm}\simeq i A_i \Psi^\pm_i,\quad A_i\equiv T^{-1}\partial_i T,\\
    &&\Psi^+=U^\dagger \partial_i U, \quad \Psi^-=U\partial_i U^\dagger,
\end{eqnarray*}
and the substitution of this expansion into $S^{(2)}$ leads to
\begin{align*}
S^{(2)}=\frac{1}{4}\sum_s\int d^2k\, \tr(\Psi^s_i \Psi^s_j)\int d^2 x\,\tr(A_iP^s A_j P^s).
\end{align*}
Using that $U^\dagger \partial_i U=-\partial_i U^\dagger U$,  the $k$-dependent part in this expression becomes $\int d^2k\, \tr(\Psi^+_i \Psi^+_j)=\int d^2k\, \tr(\Psi^-_i \Psi^-_j)=\sigma_{ij}^{(0)}$, as defined in section \ref{sec:effectiveTheory}.
Turning to the real space part, a quick calculation shows that $\sum_s \tr(A_i P^s A_jP^s)=\frac{1}{4}\tr([A_i,\tau_3][A_j,\tau_3])-\tr(A_i A_j)=\frac{1}{4}\tr(\partial_i Q \partial_j Q)-\tr(A_i A_j)$. In combination with $\sigma^{(0)}_{ij}$, the first term on the r.h.s. yields Eq.~\eqref{eq:SDiffComplete}, and the second, gives a contribution
\begin{align}
	\label{S2Superflous}
	-\frac{1}{2}\sigma_{ij}^{(0)}\tr(A_i A_j),
\end{align}
which we will see cancels against another one from the expansion of $S^{(1)}$.

\subsection{Topological action} 
\label{sub:topological_action}

 We now focus our attention on the first order term
$S^{(1)}$. Naively, one might think that a term of first order in a gradient
expansion might vanish by symmetry. On the other hand, this is a system with inbuilt
chirality. The key to understanding who wins the argument lies in an elegant trick,
first applied by Pruisken~\cite{pruisken1984} in a slightly different manner adapted to the IQH.

We start by considering the `$+$' contribution to the action $S^{(1)}$,
\begin{align*}
	S^{(1)+}&=-\frac{1}{2}\tr((T^{-1}U^\dagger T U - \mathds{1})P^+)=\cr 
	&=\frac{1}{2}\int_0^t ds\,\partial_s \tr(T^{-1}U^\dagger T UP^+),
\end{align*}
where we have introduced a continuous interpolation $s:[0,t]\to
\mathrm{U}(2),s\mapsto U(s)$, such that $U(t)=U$, and $U(0)=\mathds{1}$. (In
Pruisken's construction, the role of $U$ is taken by the Green function of electrons
in a magnetic field, and the interpolation parameter is energy running from the
bottom of the band to the Fermi energy.)

Doing the derivative, and defining $\Psi^+_s\equiv U^\dagger\partial_s U$, this can be rewritten as 
\begin{align*}
	S^{(1)+}&=\frac{1}{2}\int_0^t ds\,\tr(\left[\Psi_s, T^{-1}\right] [U^\dagger,T U] P^{+}).
\end{align*}
To derive this result we have used that  in the replica limit ${\rm tr} \bigl( T(\bx) P^{+} T^{-1}(\bx)\bigr) = R \to 0$ for each $\bx$.
The presence of two commutators implies a minimum of two extra derivatives, meaning
that each commutator individually can be evaluated at lowest order in the Moyal
expansion. Application of~\eqref{MoyalDiagonal} then gives
\begin{align*}
	S^{(1)+}&=\frac{1}{2}\int_0^t ds\,\tr(\partial_i\Psi^+_s\Psi^+_j )\,\tr( \partial_i T^{-1}  \partial_j T  P^{+}).
\end{align*}
To make further progress with this representation, we need a technical formula proven in Appendix~\ref{sec:proof_of_eq_eq:wauxformula}, 
\begin{align}
\label{eq:wAuxFormula}
	&{b}_{ij}\equiv \tr((\partial_j\Psi^+_s)\Psi^+_k )={b}_{ij}^s + {b}_{ij}^a,\\
	&{b}_{ij}^s=-\frac{1}{2}\partial_s\,\tr(\Psi^+_i \Psi_j^+),\quad {b}_{ij}^a=-\frac{1}{2}\epsilon_{ij}\tr(\Psi^+_s[\Psi^+_x ,\Psi^+_y]) ,\nonumber
\end{align}
splitting the $\bk$-part of the action into a contribution symmetric and
antisymmetric in the indices $i,j$. Concerning the former, we note $\int_0^s ds\,
{b}_{ij}^s=-\frac{1}{2}\,\tr(\Psi^+_i \Psi_j^+)=-\frac{1}{2}\,\tr(\Psi^-_i
\Psi_j^-)=\frac{1}{2}\sigma_{ij}^{(0)}$. Using that $P^++P^-=\mathds{1}$, the
symmetric contribution to $S^{(1)}$, obtained by adding the `$+$' and the `$-$', part thus assumes the form $S^{(1),s}=\frac{1}{2}\sigma_{ij}^{(0)}\,\tr(A_i A_j)$, and cancels against~\eqref{S2Superflous}.

Turning to the antisymmetric part, we note that
\begin{align*}
&\int_0^t ds\,\epsilon^{lm}\tr(\Psi^+_s\Psi^+_l \Psi^+_m)=\cr 
& =\frac{1}{4\pi^2}\int ds d^2 k \,\tr(U^\dagger \partial_s U [U^\dagger \partial_1 U,U^\dagger \partial_2 U])=\frac{1}{\pi}\Gamma[U],	
\end{align*}	
where $\Gamma[U]$ is the WZW functional of Eq.~\eqref{eq:WZWFunctional}. The corresponding `$-$' term gives the same contribution with an opposite sign. This means that the antisymmetric contribution assumes the form
\begin{align*}
 	S_\mathrm{top}=-\frac{1}{4\pi}\Gamma[U]\int d^2x\,\epsilon_{ij}\,\tr(A_i A_j \tau_3).
 \end{align*}
 Finally, a quick calculation shows that $\epsilon_{i j} \,\tr\left(A_{i} A_{j} \tau_{3}\right)=-\frac{1}{4} \epsilon_{i j} \tr\left(Q \partial_{i} Q \partial_{j}Q\right)$, which leads us to Eq.~\eqref{Stop}.

\subsection{Frequency action} 
\label{sub:frequency_action}

 We conclude this section with a quick derivation
of Eq.~\eqref{eq:FreqAction}. Retracing the steps leading to Eq.~\eqref{eq:CFTAFAI},
we note that in the presence of a finite difference, $\omega$, $U\to U e^{i \omega/2}$,
$U^\dagger \to U^\dagger e^{i\omega/2}$, the argument in the second logarithm
generalizes to $ZU^\dagger Z^\dagger U\to e^{i \omega - 2\delta}ZU^\dagger Z^\dagger
U\simeq ZU^\dagger Z^\dagger U + i \omega^+ ZZ^\dagger$, where in the second
step we assumed smallness of the decoherence parameter $\omega^+\equiv \omega+2i\delta$, and  noted that to leading order
in an expansion the non-commutativity of $Z,Z^\dagger$ with $U$ may be neglected in this term. The first order expansion of the action in $\omega^+$ then yields, $S=S_0+i \omega^+\,\tr(1+ZZ^\dagger)^{-1}ZZ^\dagger)$. With Eq.~\eqref{eq:TDef} 
and $Q=T\tau_3 T^{-1}$, a straightforward calculation shows the equivalence to Eq.~\eqref{eq:FreqAction}.

\section{Summary and discussion} 
\label{sec:summary_and_discussion}

In this paper, we have presented an analytical theory of topological quantum
criticality in the AFAI. We have seen that the structure of this theory is
essentially determined by symmetries and geometric constraints involving both the
clean part of the Floquet operator, and an effective real space field describing
phase coherent transport in the presence of disorder. On the same grounds, we have
reasoned that the finite size AFAI does not admit the definition of an integer
invariant, and that topological quantization is an emergent feature in the
thermodynamic limit. The flow towards a configuration with integer quantized
transverse response is equivalent to that in the static quantum Hall insulator.
Formally, this connection follows from the equivalence of the effective theory of the
AFAI to the Pruisken theory of the IQH. Physically, it reflects the stabilization of
edge transport via bulk localization. The distinct topological phases, $\sigma=0,1$,
are separated by a quantum phase transition which, likewise, is in the IQH
universality class. A not entirely obvious conclusion from this equivalence is that
extended Landau level center states connecting opposite surfaces of the system ---
present in the IQH, but absent in the present system --- are not essential to IQH
universality. The results discussed in this paper were obtained for an
implementation of the AFAI via a specific five step driving protocol. One may ask how
robust they are, e.g. to changes in the disorder distribution and/or changes in the
choreography of the driving protocol. While the very nature of the question excludes
a rigorous answer, it all boils down to the robustness of the AFAI quantum Hall phase
itself: how violent does a perturbation imposed on an edge mode supporting
topological thermodynamics phase have to be to destruct that mode? Since the phase is
stabilized by two robust principles, $2d$ Anderson localization, and the oriented
hopping dynamics of the clean system, we believe in a high level of protection.
Within the framework of the field theoretical construction this reflects in the well
known robustness of, say,  Wess-Zumino functionals to the presence of perturbations
at the microscopic level. However, it would take a study way beyond the present one quantitatively probe the tolerance windows of the modeling.

The geometric principles underlying the present construction carry over to other  FTI in other dimensions and symmetry classes. In fact, they can be extended beyond the class of topological insulators to describe a topologically protected Floquet \emph{metals}. These systems  have no analog in static quantum matter and will be the subject of a forthcoming publication. 

\acknowledgments 
We wish to thank Erez Berg, Netanel Lindner, and Mark Rudner for discussions. 
KWK acknowledges financial support from IBS (Project Code No.~IBS-R024-D1) and “Overseas
Research Program for Young Scientists”  through the Korea Institute for Advanced Study (KIAS).
AA and DB were funded by the Deutsche Forschungsgemeinschaft (DFG)
Projektnummer 277101999 TRR~183 (project A01/A03).
TM acknowledges financial support by Brazilian agencies CNPq and FAPERJ.

\appendix

\section{Source terms} 
\label{app:source_terms}

In the following we demonstrate that the coupling of the theory to sources as in
Eq.~\eqref{eq:SourceDef},~\eqref{eq:FieldTheorySource}, and~\eqref{eq:FieldTheorySource1} yields the disorder
averaged linear response coefficients~\eqref{eq:Sigma_H} and \eqref{eq:Sigma_xx}. The way to show this is to
retrace the steps leading to the effective action, with the sources kept in place.
Performing the source differentiation at the early steps of the construction then
yields Eqs.~\eqref{eq:Sigma_H} and \eqref{eq:Sigma_xx}.

Since $Q\to S Q S^{-1}$ and $Q=T\tau_3 T^{-1}$, the coupling of sources is equivalent
to the replacement $T\to S T\equiv T_S$. The sourced action is thus generated by
expansion of the prototype action Eq.~\eqref{eq:ProtoTypeAction} evaluated on $T_S$. From here, one may go back to the original $\psi$-representation in a sequence of operations that is straightforward yet somewhat tedious: 

We first note that the same block matrix operations that led from Eq.~\eqref{eq:CFTAFAI} to Eq.~\eqref{eq:ProtoTypeAction} show that the two terms $s=\pm$ in the latter are identical. In the present section, we find it convenient to  work with the `$-$' variant, so that the sourced action reads
\begin{widetext}
\begin{align*}
S[T_S]&= -\tr \ln (1+  T_S^{-1}(UT_S U^\dagger-T_S)P^{-})=
    \tr\ln(T)-\tr \ln (T+  (UTU^\dagger-T+(S^{-1}S^U) UTU^\dagger)P^{-}),
  \end{align*}  
where, up to the required second order in the source parameters $q_i$ of Eq.~\eqref{eq:SourceDef},
\begin{align*}
S^{-1}S^U\equiv S^{-1}(U S U^\dagger-S)=
 P^{(1)} \left(
\begin{matrix}
w&z \cr \bar z& \bar w	
\end{matrix}
\right),	
\end{align*}
where 
\begin{equation}
w\equiv -i \epsilon^{ij}q_i q^U_j - q_i q^U_i + \tfrac 12 \left( U q_i^2 U^\dagger - q_i^2\right), \qquad
z=q^U_1-i q^U_2.
\end{equation}
With the help of Eq.~\eqref{eq:TDef}, the action can then be written as
\begin{align*}
S=\tr\ln(T)-\tr\ln \left(
\begin{matrix}
	1&U ZU^\dagger+P^{(1)}(w UZU^\dagger +z)\cr
	-Z^\dagger & 1+P^{(1)}(\bar z UZU^\dagger + \bar w)
\end{matrix}
\right).
\end{align*}
We proceed by representing the second determinant as a Gaussian integral and un-doing the cft. 
On the 1st step of this procedure the fermion action is of the form  
\begin{equation}
	S_{{\rm I}}=\tr\ln(T)+\bar \psi^+\psi^++(\bar \psi^+(1+P^{(1)}w)+\bar \psi^- P^{(1)}\bar z)UZU^\dagger \psi^-
+\bar \psi^+P^{(1)}z\psi^- -\bar \psi^-Z^\dagger \psi^++\bar \psi^-(1 + P^{(1)}\bar w)\psi^-.
\end{equation}
On introducing auxiliary spinors 
\begin{equation}
\qquad \phi^- = U^\dagger \psi^-, \qquad 
\bar \phi^+ = \left(  \bar\psi^+(1 + P^{(1)} w) + \bar \psi^- P^{(1)}\bar z  \right) U, \qquad
\end{equation}
this action becomes
\begin{equation}
S_{{\rm I}}=\tr\ln(T)+\bar \psi^+ \psi^- + \bar \psi^+P^{(1)}z\psi^- +\bar \psi^-(1 + P^{(1)}\bar w)\psi^-
+  \bar \phi^+ Z \phi^- - \bar \psi^-Z^\dagger \psi^+,
 \end{equation}
and its $Z$-dependent part can be further subjected to cft, see Eq.~(\ref{eq:cftNative}). In this way we arrive at
\begin{equation}
S_{{\rm II}} = S_0 - \bar\psi^{+1} wU_{\rm d} \psi^{+1}  + \bar\psi^{-1} \bar w\psi^{-1}+\bar\psi^{+1} z\psi^{-1}-\bar \psi^{-1} \bar z U_{\rm d} \psi^{+1},
\end{equation}
where  $S_0$ is the source-free action, and the numerical superscript in $\psi^{\sigma,1}$ refers to the replica index. At this point, the differentiation in the source parameters $a_i$ sitting in 
$w$ and $z$ ---  they are expressed in terms of $q_i \equiv a_i x_i$ and  $q^U_i \equiv a_i v_i$, --- can be carried out. 
For the Hall conductivity this leads to
\begin{align}
\label{eq:sigma_xy_da1da2}
\widetilde\sigma_H &= 
\frac{1}{2i A}\left.\frac{\partial^2}{\partial a_1 \partial a_2} \right|_{a=0}\mathcal{Z}(a)\cr &=
\frac{1}{2i A}\left\langle -i \bar \psi^{+1} \epsilon^{ij}x_i v_j U_{\rm d} \psi^{+1} 
- i \bar \psi^{-1} \epsilon^{ij} x_i v_j  \psi^{-1}- i\bar \psi^{+1} v_1 \psi^{-1}\bar \psi^{-1} v_2 U_{\rm d} \psi^{+1}
+ i \bar \psi^{+1} v_2 \psi^{-1}\bar \psi^{-1}v_1 U_{\rm d} \psi^{+1}\right\rangle\cr 
&= \frac{1}{2 A}\Bigl( \epsilon^{ij}\tr(  x_i v_j)+ \tr( v_1 G^-  v_2 G^+-   v_2 G^- v_1 G^+) \Bigr),
\end{align}
where in the last step we used that $\langle G^-\rangle =1$ and $\langle G^+U_{\rm d}\rangle=0$ 
due to the random fluctuations of $U_{\rm d}$, while 
$\epsilon^{ij} \langle v_i G^- v_j U_{\rm d} G^+ \rangle = \epsilon^{ij} \langle v_i G^- v_j G^+ \rangle$. The latter can be seen by
representing $U_{\rm d} = 1- (G^{+})^{-1}$ in the limit $\delta \to 0^+$.
The final result~(\ref{eq:sigma_xy_da1da2}) contains the two terms entering the correlation function~\eqref{eq:Sigma_H}. 
Similarly, for the longitudinal conductivity one derives
\begin{align}
\label{eq:sigma_xx_da1da1}
\widetilde\sigma_{11} &= 
\frac{1}{2A}\left.\frac{\partial^2}{\partial a_1^2} \right|_{a=0}\mathcal{Z}(a)\cr &= 
\frac{1}{A}\left\langle - \bar \psi^{+1} x_1 v_1 U_{\rm d} \psi^{+1} + \tfrac 12  \bar \psi^{+1} [U_{\rm d}, x_1^2] \psi^{+1}
 + \bar \psi^{-1} x_1 v_1  \psi^{-1} - \tfrac 12 \bar \psi^{-1} ( U_{\rm d} x_1^2 U_{\rm d}^\dagger - x_1^2)  \psi^{-1} 
- \bar \psi^{+1} v_1 \psi^{-1}\bar \psi^{-1} v_1 U_{\rm d} \psi^{+1}
\right\rangle\cr
&= \frac{1}{A}\Bigl(   - \tr(  x_1 v_1) + \tr( v_1 G^-  v_1 U_{\rm d} G^+) \Bigl)  = 
\frac{1}{A}\Bigl(   - \frac 12  \tr(  v_1^2) + \tr( v_1 G^-  v_1  G^+) \Bigl).
\end{align}
\end{widetext}
To get the very last result above we have once again employed $U_{\rm d} = 1- (G^{+})^{-1}$ and used the relation
$- \tr( x_1  v_1) = \tfrac 12 \tr(  v_1^2)$. In its final form the result~(\ref{eq:sigma_xx_da1da1}) equals to
the correlation function~(\ref{eq:Sigma_xx}).
To summarize, the above calculations establish the equality $\widetilde\sigma_H=\sigma_H(L)$ and $\widetilde\sigma_{11} = \sigma_{11}(L)$  of the running coupling constants of the field theory at scale $L$ and the microscopically defined response functions. 

\section{Proof of the equivalence $\sigma^{\rm II} = \sigma_H^{(0)}$} 
\label{sec:proof_of_equiv}
The proof of this equivalence is very similar to the derivation of the topological action in Sec.~\ref{sub:topological_action}.
On introducing the interpolation $U(s)$ as it was done before and using $\Psi_s^- \equiv U \partial_s U^\dagger$ one has
\begin{eqnarray}
\sigma^{\rm II} &=& \frac {\epsilon_{ij}}{2 A} \int_0^{{t}} d{s} \, \partial_s {\rm tr} (x_i U x_j U^\dagger)  \nonumber\\
&=& \frac {\epsilon_{ij}}{2 A} \int_0^{{t}} d{s} \, {\rm tr} ( x_i U x_j U^\dagger \Psi^-_{{s}}  - x_i \Psi^-_{{s}} U x_j U^\dagger) \nonumber\\ 
&=& \frac {\epsilon_{ij}}{2 A} \int_0^{{t}} d{s} \, {\rm tr} \left(  [\Psi^-_{{s}}, x_i] U x_j U^\dagger\right).
\end{eqnarray}
At this stage we use the exact relations
\begin{equation}
U x_j U^\dagger -  x_j = i U \partial_j U^\dagger, \qquad 
[\Psi_s^-, x_i] = - i \partial_i \Psi^-_s,
\end{equation}
and an obvious identity ${\rm tr} ( [\Psi_{{s}}^-, x_i] x_j) \epsilon_{ij} = 0$. Taking them into account by means of the following chain of  transformations,
\begin{eqnarray}
\sigma^{\rm II} &=& \frac {\epsilon_{ij}}{2 A} \int_0^{{t}} d{s} \, {\rm tr} \left( \partial_i \Psi_s^- U \partial_j U^\dagger \right)  
\nonumber \\
&{=}& -\frac{\epsilon_{ij}}{8 \pi^2}\int ds  d^2 k \,{\rm tr} 
\left( \Psi_s^-  \partial_i U \partial_j U^\dagger \right) \nonumber \\
&  = & \frac{\epsilon_{ij}}{8 \pi^2}\int ds  d^2 k \,{\rm tr} 
\left( \Psi_s^-  \Psi_i^- \Psi_j^- \right) = \frac{1}{2\pi} \Gamma[U] = \sigma_H^{(0)} \nonumber,
\end{eqnarray}
one accomplishes the proof, where in the second line we have used an integration by parts in momentum space.\newline

\section{Proof of Eq.~\eqref{eq:wAuxFormula}} 
\label{sec:proof_of_eq_eq:wauxformula}

To prove Eq.~\eqref{eq:wAuxFormula}, we first note that the trace in ${b}_{ij}\equiv
\tr((\partial_j\Psi^+_s)\Psi^+_k )=\tr(\partial_j(U^\dagger \partial_s U)U^\dagger \partial_k U)$ implies an integration over momenta, $\bk$, with
periodic boundary conditions. We may thus rearrange momenta via integrations by
parts. Applying this freedom to the $j$-derivative, we obtain
\begin{align*}
	{b}_{ij}&=\frac{1}{2}\tr(\partial_i(U^\dagger \partial_s U)(U^\dagger \partial_j U-\partial_j U^\dagger U))=\cr
&=	-\frac{1}{2}\tr(U^\dagger \partial_s U (\partial_i U^\dagger \partial_j U-\partial_j U^\dagger \partial_i U + \cr 
&\hspace{2cm}+U^\dagger \partial^2_{ij}U-\partial^2_{ij}U^\dagger U))=\\
	&=\frac{1}{2}\tr(\Psi_s^+[\Psi_i^+,\Psi_j^+]+ \partial_s U^\dagger \partial^2_{ij} U+\partial_s U \partial^{2}_{ij}U^\dagger).
\end{align*}
The first term in this expression defines the anti-symmetric contribution to Eq.~\eqref{eq:wAuxFormula}. A double partial integration in $k_{i,j}$ brings the second into the form, $\frac{1}{2}\partial_s\,\tr(U^\dagger \partial^2_{ij}U)=-\frac{1}{2}\partial_s\,\tr(\Psi^+_i \Psi_j^+)$, which is the symmetric contribution.

\section{Numerical details} 
\label{sec:numerics}

To compute the Hall, $\widetilde\sigma_H$, and longitudinal, $\widetilde\sigma_{11}$, conductivities for a specific disorder realization in Fig.~\ref{fig:sigma_num}, we have used an enlarged unit cell of lattice size $\rm L_1\times L_2$ with a twisted boundary condition across the unit cell boundary by adding phase $\theta_{1,2}$ in the hopping amplitudes:
\begin{eqnarray}
U^{\bold{\theta}}_{\bx\by}&=&U_{\bx\by}e^{i\theta_1\delta_{x_1-y_1,\rm L_1-1}+i\theta_2\delta_{x_2-y_2,\rm L_2-1}} \nonumber \\
&&\times e^{-i\theta_1\delta_{x_1-y_1,-\rm L_1+1}-i\theta_2\delta_{x_2-y_2,-\rm L_2+1}},
\end{eqnarray}
where $\bx=(x_1,x_2)$, $\by=(y_1,y_2)$, and $U_{\bx\by}$ is a matrix element of the clean 
Floquet operator in real space. Notice that our model contains nearest neighbor hopping only. The velocity operators are then computed through $v_i=U^{\bold{\theta}\dagger}\partial_{\theta_i}U^{\bold{\theta}}$, 
which in fact is independent both of $\theta_{1,2}$ and disorder. 
By taking a set of eigenvalues and eigenvectors of $U^{\bold{\theta}}$,  $\sigma^I$ in \eqref{eq:sigma_12I_final} and $\widetilde\sigma_{ii}$ in \eqref{eq:sigma_ii} are computed for a specific twisted boundary $(\theta_1,\theta_2)$. Lastly, the conductivities are averaged over  $\theta_{1,2}\in[0,2\pi]$. 

\section{Hall conductivity} 
\label{sec:sigma_Hall}

In this appendix we derive equivalent representation for the Hall response
coefficient, $\tilde\sigma_H = \sigma^{\rm I} + \sigma^{\rm II}$, which in the
localized regime will enable us to relate $\tilde\sigma_H$ to the number of topological
chiral edge modes. We start from Eqs.~(\ref{eq:sigma_12I_final}) and
(\ref{eq:sigma_II_WZW}) and transform $\sigma^{\rm II}$ in a different way as
compared to Appendix~\ref{sec:proof_of_equiv}. To this end, consider a cylindrical
geometry of a width $L_1$ and a circumference $L_2$ implying periodic boundary
conditions along $x_2$-direction. A cylinder has two boundaries where delocalized
chiral edge states may  propagate. In addition to the velocity operator $v_2 = U x_2
U^\dagger -x_2$ we introduce another one,
$\bar v_2 = - U^\dagger x_2 U+ x_2$.
With the help of the latter 
\begin{eqnarray}
{\rm tr} ( v_1 x_2) &=& {\rm tr} ( U x_1 U^\dagger - x_1) x_2 \\
 &=& 
{\rm tr} \,x_1( U^\dagger x_2 U -  x_2) = - {\rm tr} ( x_1 \bar v_2 ), \nonumber
\end{eqnarray}
and therefore
\begin{equation}
\sigma_{12}^{\rm II} = \frac 12 {\rm tr} \, x_1 (v_2 + \bar v_2). 
\end{equation}
To proceed, we first analyze the off-diagonal terms of $\sigma_{12}^{\rm II}$ in the
eigenbasis of $U$.  Denoting the off-diagonal contribution as $\widetilde
\sigma_{12}^{\rm II}$, one can write for the matrix elements of velocity operators
$v_2$ (below $m \neq n$),
\begin{eqnarray}
\langle n |  v_2 |m\rangle &=& \langle n |  x_2 |m\rangle (e^{i(\epsilon_m - \epsilon_n)}-1), \\
\langle n |  \bar v_2 |m\rangle &=& \langle n |  x_2 |m\rangle (1 - e^{-i(\epsilon_m - \epsilon_n)}),
\end{eqnarray}
which yields a nice cancellation,
\begin{eqnarray}
\widetilde \sigma_{12}^{\rm II}  &=& 
\frac i A \sum_{m\neq n} \langle m |  x_1 |n\rangle \langle n | x_2 | m \rangle \sin (\epsilon_m - \epsilon_n)   \\
&=&
\frac {i}{2A} \sum_{m\neq n}\langle m | v_ 1 |n\rangle \langle n | v_2 | m \rangle \cot \left(\frac{\epsilon_m - \epsilon_n}{2}\right) 
= -\sigma_{12}^{\rm I}, \nonumber
\end{eqnarray}
where in the last line Eq.~(\ref{eq:sigma_12I_final}) was used.
Hence, only diagonal terms contribute to the final result and
\begin{equation}
\widetilde\sigma_H  =  \frac{1}{2 A}
\sum_{m} \langle m | x_1 | m \rangle \langle m | v_2 + \bar v_2 | m \rangle.
\end{equation}
At this point, a subtlety comes into play --- the diagonal matrix elements of velocity operators $v_2$ and $\bar v_2$
can be only defined without a reference to the position operator $x_2$.  In the clean limit and infinite circumference, 
$L_2  \to \infty$,  one can set $x_2 = i \partial_{k_2}$ (where $k_2 \in [-\pi,\pi]$ is the Bloch momentum) so that
\begin{eqnarray}
\langle m | v_2| m \rangle  &=&  
\langle m | U  ( i \partial_{k_2} U^\dagger )| m \rangle, \nonumber \\ 
\langle m | \bar v_2| m \rangle &=&  
- \langle m | U^\dagger ( i \partial_{k_2} U)| m \rangle. 
\end{eqnarray}
Provided the system is homogeneous in $x_2$-direction, the momentum $k_2$ is conserved and the eigenstate
has the form $| m \rangle \equiv |l , k_y \rangle$, where $l$ is a discrete quantum number stemming from the quantization
along $x_1$-direction. For this configuration we find
\begin{equation}
\label{eq:Zak_xk}
\widetilde\sigma_H  =  \frac{1}{L_1}
\sum_{l} \int_0^{2\pi} \frac{d k_2}{2\pi} \langle l, k_2 | x_1 | l, k_2 \rangle \frac{\partial \epsilon_l (k_2)}{\partial k_2}.
\end{equation}

The above result can be now extended to the general situation when $L_2$ is finite and 
the periodically driven Floquet system includes spatial disorder. 
In this case there is no continuous momentum variable $k_2$ to differentiate. However,
one can define the Floquet operator $\hat U(\theta_2)$ for a system with twisted boundary conditions in $x_2$-direction
which has a flux-dependent spectrum of quasi-energies $\epsilon_m(\theta_2)$, see Appendix~\ref{sec:numerics}. 
In physical terms it corresponds to the flux $\theta_2$ of a magnetic field threading a cylinder along the axis $x_1$.
Then the above result~(\ref{eq:Zak_xk}) should be generalized as
\begin{equation}
\label{eq:Zak_x}
\widetilde\sigma_H = \frac{1}{L_1}
\sum_{m}\int_0^{2\pi} \frac{d\theta_2}{2\pi}  \langle m,\theta_2 | x_1 |m, \theta_2 \rangle  
\frac{\partial \epsilon_m (\theta_2)}{\partial \theta_2}.
\end{equation}
To justify this expression, we refer to a well known trick: one considers an auxiliary periodic in $x_2$-direction 
Floquet system composed of infinite number of supercells of length $L_2$ sharing the same realization of disorder, 
and then appeals to the previous Eq.~(\ref{eq:Zak_xk}).

The relation~(\ref{eq:Zak_x}) has a transparent meaning in the localized phase where,
as we show below, it equals to the number of chiral edge modes, $n_{\rm edge}$.
Indeed, the states localized in the bulk are not sensitive to a variation of
$\theta_2$ (for them $ \partial \epsilon_m (\theta_2) / \partial \theta_2 \to 0$) and
thus do not contribute to (\ref{eq:Zak_x}). On other hand, for the set of right/left
boundary states, which we denote as $|b^\pm, \theta_2 \rangle$ where $0 \leq b^\pm <
{\cal N}$,  the matrix elements $\langle b^\pm ,\theta_2 | x_1 |b^\pm , \theta_2
\rangle  \to \pm L_1/2$ and group velocities are non-zero, $\partial \epsilon_{b}^\pm
(\theta_2) / \partial \theta_2 \gtrless 0$. Importantly, these states are subjected
to the spectral flow, i.e.
\begin{equation}
\label{eq:Spectral_flow}
\epsilon_{b}^\pm(\pi-0) \to \epsilon_{b+1}^\pm(-\pi + 0), \quad b=0, \dots, \,{\cal N}-1. 
\end{equation}
and thereby form either a single or a number of chiral edge modes, cf. Fig.~3 in Ref.~\onlinecite{titum2016}.
Note that the number of chiral edge states, $n_{\rm edge}$, is smaller than the number of edge states, ${\cal N}$,  
since few $b$'s will typically combine to create a chiral mode. Hence, for $\widetilde \sigma_H$ we find
a limiting expression valid in the Anderson localized phase,
\begin{equation}
\label{eq:Zak_x_loc}
\widetilde\sigma_H \to \frac 12
\sum_{b=0}^{{\cal N}-1} \int_0^{2\pi} \frac{d\theta_2}{2\pi}  
\left(   
\frac{\partial \epsilon_b^+ (\theta_2)}{\partial \theta_2} - \frac{\partial \epsilon_b^- (\theta_2)}{\partial \theta_2} 
\right).
\end{equation}
Because of the spectral flow~(\ref{eq:Spectral_flow}), the above relation shows 
how many times the quasi-energy spectrum winds around a cirlce (the full Bloch zone) $[-\pi, \pi]$ as the flux $\theta_2$ varies 
from 0 to $2\pi$ and in this way $\widetilde \sigma_H$ counts 
the number of chiral states $n_{\rm edge}$. The analysis in this Appendix is in line
with the previous work by Titum {\rm et al.}\cite{titum2016}, see in particular their discussion in section III.B.
 
\bibliography{bibliography}

\end{document}